\documentclass[11pt]{article}
\usepackage{jheppub} 
\makeatletter
\gdef\@fpheader{}
\makeatother
\usepackage[normalem]{ulem}
\usepackage{amsmath,amsfonts,amssymb,bbm}
\usepackage{units}
\usepackage{hyperref}
\usepackage{graphicx}
\usepackage{feynmp}
\usepackage{makecell}
\usepackage{bm}
\usepackage{booktabs,multirow}
\usepackage[usenames,table]{xcolor} 
\usepackage{slashed}                
\usepackage{parskip}                
\usepackage{array}                  
\usepackage[section]{placeins}      
\usepackage{enumitem}               
\usepackage{moresize}               
\usepackage[utf8]{inputenc}         
\usepackage{fancyhdr}               

\newcommand{\subref}[2]{\hyperref[#1]{\ref*{#1}#2}}
\def\code#1{\texttt{#1}}

\renewcommand{\to}{\rightarrow}

\def\et{\ifmmode \epsilon (\tau) \else $ \epsilon (\tau)$ \fi}
\def\ejt{\ifmmode \epsilon ( j \to \tau) \else $ \epsilon (j \to \tau) $ \fi}

\subheader{\rm IFT-UAM/CSIC-25-5 \par
\rm COMETA-2025-03}

\title{\texorpdfstring{\textbf{Exotic $\bm{h \to Z a}$ Higgs decays into $\bm{\tau}$ leptons}}{Exotic h → Z a Higgs decays into tau leptons}}

\author[a]{M. Cepeda,}
\author[b,c]{J.~M. No,}
\author[d]{C.~Ramos,}
\author[b,c]{R. M.~Sand\'a Seoane,}
\author[e]{J.~Zurita}
\affiliation[a]{Centro de Investigaciones Energ\'eticas, Medioambientales y Tecnol\'ogicas, E-28040, Madrid, Spain}
\affiliation[b]{Departamento de F\'isica Te\'orica, Universidad Aut\'onoma de Madrid, Cantoblanco,\\ E-28049, Madrid, Spain}
\affiliation[c]{Instituto de F\'isica Te\'orica IFT-UAM/CSIC, 
Cantoblanco, E-28049, Madrid, Spain}
\affiliation[d]{Centro de Ci\^encias Naturais e Humanas, Universidade Federal do ABC, Santo Andr\'e, 09210-580 SP, Brazil}
\affiliation[e]{Instituto de F\'{\i}sica Corpuscular, CSIC-Universitat de Val\`encia, E-46980, Paterna, Valencia, Spain}
\emailAdd{maria.cepeda@cern.ch}
\emailAdd{josemiguel.no@uam.es}
\emailAdd{ramos.camila@ufabc.edu.br}
\emailAdd{rosa.sanda@uam.es}
\emailAdd{jzurita@ific.uv.es}

\abstract{
Exotic Higgs decays are among the most promising areas to be explored at the High-Luminosity LHC, given the unprecedentedly large amount $(\sim 3 \times 10^8)$ of 125 GeV 
Higgs bosons that will be produced. In this context, we propose 
a new search channel for which the Higgs boson decays to a (leptonically decaying) $Z$ boson and a light BSM pseudoscalar $a$, which subsequently decays to a pair of $\tau$-leptons ($h \to Z a \to \ell\ell \tau\tau $). 
After performing a validation of existing ATLAS and CMS exotic Higgs decay searches in related channels, we analyze the HL-LHC projected sensitivity of our $a\to \tau\tau$ search, targeting the kinematic region where the exotic Higgs decay is two-body. We are able to probe pseudoscalar masses $m_a \in [5,\, 33]$ GeV by leveraging both leptonic and hadronic $\tau$ decays, and establish model-independent 95\% C.L. sensitivity projections on the branching fraction ${\rm BR}(h \to Z a) \times {\rm BR}(a \to \tau\tau)$. These $a\to \tau\tau$ projections yield a competitive 
probe of light pseudoscalars, which depending on the model can become significantly more sensitive than projections from existing experimental searches in $a \to \mu\mu$ and $a \to \gamma\gamma$ final states.
Finally, we explore the potential of our search to probe an Axion-Like-Particle (ALP) solution to the muon $(g-2)$ anomaly (when taken face-value), finding that our proposed $h\to Z a$, $a\to\tau\tau$ search can provide valuable constraints on such ALP scenario, in complementarity with existing $h\to Z a$, $a \to \gamma\gamma$ experimental searches.

}

\begin{document}
\maketitle

\section{Introduction}
\label{sec:1}

In the quest to look for 
beyond the Standard Model (BSM) physics, undoubtedly the 125 GeV Higgs boson $h$ plays a unique role. On the one hand, once its mass is known, all SM Higgs couplings -- and hence its production and decay modes -- are forecast with high precision~\cite{LHCHiggsCrossSectionWorkingGroup:2016ypw}, resulting in a sizable sample of $h$ particles ($\sim 3 \times 10^8$) to be produced at the High-Luminosity phase of the LHC (HL-LHC). On the other hand, from a theoretical perspective, the SM Higgs doublet $H$ can lead to several renormalizable interactions with BSM states when such new particles are added to the SM.\footnote{A prime example of this is the so-called ``Higgs-portal'' scenario~\cite{Patt:2006fw}.} When these new particles are lighter than $h$, new Higgs boson decay channels, not present in the SM, may appear. These final states are known as \emph{exotic Higgs decays} (see e.g.~\cite{Curtin:2013fra,Cepeda:2021rql} for reviews), and hence the SM Higgs boson -- as a necessary ingredient in their collider production -- acts as a unique window into BSM scenarios including ``light''  -- below the electroweak (EW) scale -- BSM particles.

It is worth noting that in BSM scenarios 
one can have both \textit{i)} Higgs boson couplings to new states, as well as \textit{ii)} modifications in the expected $h$ interactions w.r.t. their SM values. For the sake of simplicity and given that the existing data on SM Higgs couplings to matter fields strongly favours a SM-like coupling pattern~\cite{ATLAS:2024fkg,CMS:2022dwd}, we will consider in what follows possible new decay channels of $h$ as our only phenomenologically relevant BSM effects.\footnote{Hence, in our analysis we will employ state-of-art predictions for the SM Higgs production cross-sections~\cite{Anastasiou:2016cez} and for its partial width into SM final states~\cite{higgsbr}.}
Given the expected number of SM Higgs bosons to be produced at the HL-LHC, one can naively expect, depending on the specific channel under consideration, to be able to probe \emph{exotic branching fractions} ${\rm BR} (h \to \rm{BSM})$ down to $10^{-4} - 10^{-5}$. In contrast, a global fit to all Higgs data is currently only able to constrain a non-SM branching fraction of about 12\%~\cite{ATLAS:2024fkg,CMS:2022dwd}, which would be lowered down to 4\% at the HL-LHC~\cite{deBlas:2019rxi}.

Most of the considered theoretical scenarios yielding exotic Higgs decays focus on $h$ decays into new scalars or pseudoscalars via $h\to ss$ or $h\to a a$
(see e.g.~\cite{Chang:2008cw,Curtin:2015fna,Kozaczuk:2019pet,Carena:2022yvx,Robens:2022erq} and references therein), 
yet decays into new fermions, such as Heavy Neutral Leptons~\cite{Caputo:2017pit,Thor:2023nzu} or those belonging to a strongly-interacting dark sector (e.g. giving rise to emerging jets)~\cite{Born:2023vll,Carrasco:2023loy,Cheng:2024aco}, as well as decays into dark photons~\cite{Curtin:2014cca}, have also been investigated. 
In this work, we focus on the decay $h\to Z a$ -- with $a$ a BSM pseudoscalar particle -- , which has so far received little attention in the literature (see however~\cite{Bauer:2017ris,Aguilar-Saavedra:2022xrb,Cheung:2024kml,Shan:2024pcc}). Such decay, kinematically open for $m_a \leq m_h - m_Z \simeq 33$ GeV, is phenomenologically viable in several theoretical scenarios. We will mention for concreteness the case where $a$ is an Axion-Like Particle (ALP)~\cite{Bauer:2017ris,Brivio:2017ije}, as well as the case of extended Higgs sectors with new pseudoscalar particles -- e.g. the Two-Higgs-Doublet Model plus pseudoscalar (2HDM$+a$) model~\cite{Ipek:2014gua,No:2015xqa,Goncalves:2016iyg,Bauer:2017ota}, a widely used benchmark scenario in LHC searches for Dark Matter~\cite{LHCDarkMatterWorkingGroup:2018ufk} -- .

Experimental searches at the LHC for the $h\to Z a$ exotic Higgs decay by the ATLAS and CMS Collaborations do exist, targeting leptonic decays\footnote{Other experimental searches for exotic Higgs decays into a four-lepton final state via $h \to X X \to 4\ell$ -- with $X$ an intermediate resonance -- also exist (see e.g.~\cite{CMS:2018jid}). Yet, these do not involve the production of a $Z$ boson from the Higgs decay.} $a \to \ell \ell$ ($\ell = e, \mu$)~\cite{ATLAS:2018coo,ATLAS:2021ldb,CMS:2021pcy} or decays into a pair of photons $a \to \gamma \gamma$~\cite{ATLAS:2023etl}. While both these final states are very clean at the LHC, they are not expected to be among the dominant decay modes of the pseudoscalar $a$ in concrete BSM models, except when the pseudoscalar mass $m_a$ is below the production threshold of other decay modes such as $a \to b \bar{b}$, $a\to \tau\tau$ or $a \to g g$. Here 
we target the cleanest among such would-be dominant pseudoscalar decay modes: $a \to \tau\tau$~\footnote{See~\cite{Alda:2024cxn} for a recent phenomenological study of $a$ particles coupling (almost) exclusively to $\tau$ pairs.}. Such channel 
yields a promising opportunity to enhance the LHC sensitivity to BSM models 
w.r.t.~current exotic Higgs decay searches. Specifically, for extended Higgs sectors or ALPs, the pseudoscalar $a$ will generally couple to the SM fermions proportionally to their mass. Given the hierarchy of SM fermion Yukawa couplings, one expects
\begin{equation}
\frac{\text{BR}(a\to \mu\mu)}{\text{BR}(a\to \tau\tau)} = \frac{m_\mu^2}{m_\tau^2} \simeq 3.6 \times 10^{-3}  \, ,
\label{eq:BR_Hierarchy}
\end{equation}
as long as the decay $a\to \tau\tau$ is kinematically open -- for $m_a \gtrsim 3.5$ GeV -- . Such large ratio between fermionic branching fractions can easily compensate for the cleaner decay into muon pairs when comparing the sensitivities of searches in both final states.\footnote{We also stress that in other BSM scenarios, e.g. if the new particle is a gauge boson $Z'$ rather than a pseudoscalar, the hierarchy between branching fractions~\eqref{eq:BR_Hierarchy} is not expected and final states with muons will in general be much more sensitive.} In fact, searches for exotic Higgs decays in the same final state that we advocate for in this work exist, i.e. $h\to X X \to \mu\mu \tau \tau$~\cite{ATLAS:2015unc,CMS:2018qvj,CMS:2020ffa} (of course the intermediate state is very different, and thus both the search analysis and the model interpretation differ), and they yield superior sensitivity to the corresponding leptonic searches $h\to X X \to \mu\mu \mu \mu$~\cite{CMS:2018jid} when interpreted in terms of concrete BSM models~\cite{Cepeda:2021rql}.

The current manuscript is organized as follows. As a validation of our simulation setup, we start by reproducing the LHC experimental searches for $h \to Z a \to \ell\ell \mu \mu$ (same intermediate state in the exotic Higgs decay) in Section~\ref{sec:AppI} and  $h \to a a \to \mu\mu \tau \tau$ (same final state in the exotic Higgs decay) in Section~\ref{sec:AppII}. In each case, we choose the most detailed experimental search that allows us to validate our analysis, respectively ATLAS~\cite{ATLAS:2021ldb} and CMS~\cite{CMS:2018qvj}. In Section~\ref{sec:experiment} we discuss in detail the prognosis of our proposed  $h \to Z a \to \ell\ell \tau \tau$ search at the HL-LHC, and present model-independent results, which can be readily applied to concrete
new physics scenarios. We exemplify the impact of this search (and the flexibility of its reinterpretation) by first comparing its sensitivity with the one obtained from $a \to \mu^+ \mu^-$ and $a\to \gamma \gamma$ final states. We then focus in Section~\ref{sec:models} on a concrete ALP scenario that aims to explain the observed $(g-2)_{\mu}$ anomaly, and show that our proposed search is complementary to other probes in accessing the $(g-2)_{\mu}$ parameter space favoured by experimental data (if the anomaly is taken face-value). Finally, we present our conclusions in Section~\ref{sec:conclusions}.

\section{\texorpdfstring{\textbf{ATLAS $\bm{h \to Z a \to \ell \ell \mu \mu}$ search validation (same intermediate state)}}{ATLAS h → Z a → llμμ search validation (same intermediate state)}}
\label{sec:AppI}

The search for exotic Higgs decay $h \to Z X$ -- with $X$ a light BSM particle -- in the final state of four light leptons ($\ell = e,\mu$), which yields an extremely clean BSM signal, has been performed by the ATLAS Collaboration~\cite{ATLAS:2018coo,ATLAS:2021ldb},  focusing primarily on a dark photon $Z_D$ interpretation ($h \to Z Z_D$). The first search~\cite{ATLAS:2018coo} was performed for $\sqrt{s} = 13$ TeV LHC data with 36.1 fb$^{-1}$ of integrated luminosity, and was subsequently reinterpreted by Bharucha et al~\cite{Brooijmans:2020yij} in the context of an ALP model, i.e. $h \to Z a \to \ell \ell \mu \mu$. The most recent ATLAS analysis~\cite{ATLAS:2021ldb} is carried out for LHC $13$ TeV and an integrated luminosity of 139 fb$^{-1}$. It does include the ALP interpretation, for which it targets the mass range 15 GeV $< m_{a} < 30$ GeV. We here perform our own re-derivation of the results of the ATLAS analysis~\cite{ATLAS:2021ldb}, which helps us calibrate our analysis tools. 

The ATLAS search selects final-state events with a \textit{quadruplet}, consisting of two pairs of same-flavour opposite-sign (SFOS) light leptons ($\ell = e,\mu$). Baseline electrons (muons) have momentum reconstruction thresholds $p_{T} > 7$ GeV ($p_{T} > 5$ GeV), and the three highest-$p_{T}$ leptons in the event must satisfy $p_T(\ell_1) > 20$ GeV, $p_T(\ell_2) > 15$ GeV, $p_T(\ell_3) > 10$ GeV, respectively.
All possible pairs of same-flavour leptons in the event must satisfy $\Delta R (\ell, \ell^{'}) > 0.1$, while different-flavour leptons must satisfy $\Delta R (\ell, \ell^{'}) > 0.2$. The invariant masses of the two lepton pairs are denoted $m_{12}$ and $m_{34}$, with $m_{12}$ the closest to the $Z$-boson mass, i.e. $|m_{12} - m_Z | < | m_{34} - m_Z |$. The mass windows 50 GeV $< m_{12} < 106$ GeV and 12 GeV $< m_{34} <$~115 GeV are imposed. For $4\mu$ and $4e$ events, the invariant masses corresponding to the alternative possible lepton pairings, $m_{14}$ and $m_{23}$, must satisfy $m_{14}, m_{23} > 5 $ GeV (to remove $J/\Psi$ contributions). When an event has multiple possible lepton quadruplets\footnote{Either because, being a $4\mu$ or $4e$ event, it admits two possible SFOS lepton pairings, or because there are more than four reconstructed light leptons in the final state of the event.} satisfying the above requirements, the quadruplet with smallest $|m_{12} - m_Z |$ is chosen. Finally, the event selection requires the invariant mass of the quadruplet $m_{4\ell}$ to be consistent with the decay of a 125 GeV Higgs boson, 115 GeV $ < m_{4\ell} < $ 130 GeV.

Before we continue the validation of the analysis, it is worth stressing that 
the broad allowed range for $m_{12}$ -- which is supposed to reconstruct the di-leptonic $Z$ boson decay, and thus one could expect it to be much narrower, e.g. $m_{12} \in [81,\, 101]$ GeV -- is justified from the fact that the $h \to Z X$ ATLAS analysis~\cite{ATLAS:2021ldb} targets the mass range $m_{X} \in [15, 55]$ GeV, such that for the high-end of the $m_X$ range the $Z$ boson may be off-shell, i.e. $h \to Z^* a \to \ell \ell \mu\mu$.

To generate signal events, we use a modified version of the Universal FeynRules Output (UFO)~\cite{Degrande:2011ua} model describing an effective field theory (EFT) extension of the SM via an ALP developed in~\cite{Bauer:2017ris}. As our signal we then consider 
$ p p \to h \to Z a$, with the subsequent decays $Z \to \ell \ell$, $a \to \ell' \ell'$. We stress that since $\text{BR}(a\to \mu \mu)/\text{BR}(a\to e e) = m_{\mu}^2/m_e^2 \simeq 4 \times 10^4$, our signal samples are almost exclusively composed of $a \to \mu \mu$ decays.   
These signal samples are generated with 
{\sc MadGraph5\_aMC@NLO}~\cite{Alwall:2014hca} and interfaced to {\sc Pythia8}~\cite{Sjostrand:2014zea} for parton shower and hadronization, and then to {\sc Delphes}~\cite{deFavereau:2013fsa} for detector simulation. We use a modified ATLAS detector card where the transverse momentum thresholds for reconstruction of electrons and muons are respectively lowered (w.r.t. to the default $p_{T} > 10$ GeV threshold) to $p_{T} > 7$ GeV and $p_{T} > 5$ GeV, to match the baseline object reconstruction of the ATLAS analysis~\cite{ATLAS:2021ldb}.

After performing the event selection discussed above, we aim to recover the analysis efficiencies $\epsilon_c$ reported by ATLAS, defined as the fraction of events passing the fiducial selection (using generator-level quantities) that also pass the full event selection (using reconstructed quantities).\footnote{Specifically, we consider the fraction of simulated events after {\sc Pythia8} (.lhe file) that pass the event selection criteria and the corresponding fraction of reconstructed events with {\sc Delphes} (.lhco file) that pass those criteria, and obtain the efficiency as their ratio. We have also set both muon and electron reconstruction efficiencies in our {\sc Delphes} detector card to 1.0, for better agreement with ATLAS.} We obtain simulated efficiencies for the $4\mu$ and $2e\,2\mu$ final states of the order of 
46\%, largely independent of $m_a$. The original ATLAS reference reports efficiencies in the range 
$\epsilon_c \sim$ 56\% - 64\% with a slight increase towards larger values of $m_a$. 
Our mild (approximately $20\%$) discrepancy with ATLAS is likely due to differences in the modeling of the reconstructed leptons at very low transverse momentum in {\sc Delphes} compared to the full reconstruction performed by ATLAS in their data analysis. Additionally, we have compared the acceptance of our ALP signal model with the dark photon model of ATLAS. Acceptance values grow 
with $m_a$ within the range 40\% – 60\% and are found to be in good agreement with the ATLAS analysis, differing by at most 9\% across the full mass range despite the fact that our signal models are not identical.

\begin{figure}[h]
  \centering
  \includegraphics[width=0.7\textwidth]{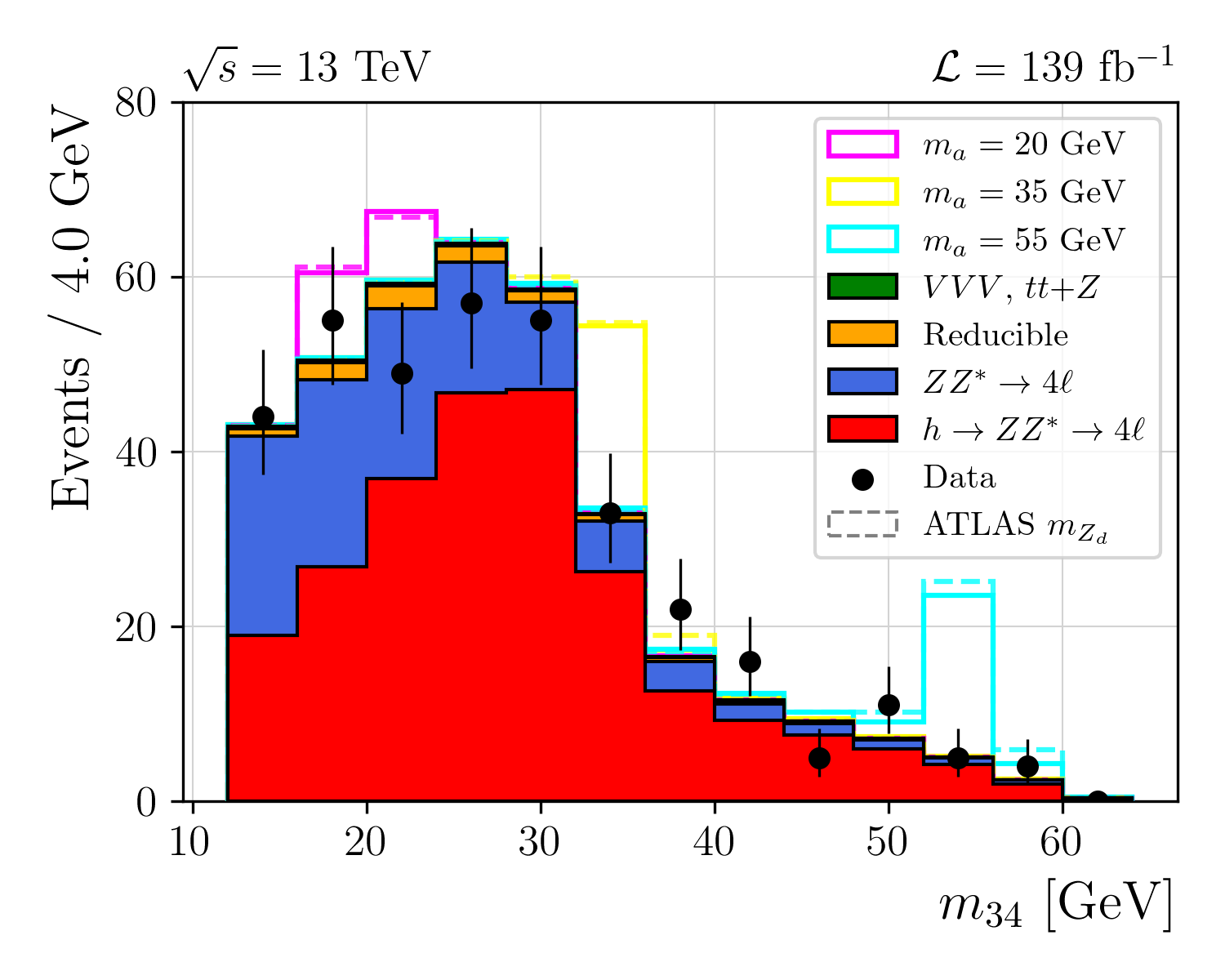}
  
  \vspace{-5mm}
  
  \caption{$m_{34}$ invariant mass distribution for reconstructed events after event selection for the SM backgrounds (filled) -- as extracted from the ATLAS analysis~\cite{ATLAS:2021ldb} -- and our simulated signal (solid lines) for masses $m_a = \{20,\, 35,\, 55\}$ GeV. We also show the signal predictions (dashed lines) for the dark-photon model used in~\cite{ATLAS:2021ldb}, as well as the observed data from the ATLAS search (black points).}
\label{fig:HiggsZa:ATLAS_efficiencies_m34_1}
\end{figure}

The $m_{34}$ distribution for the reconstructed signal events -- for benchmark mass values $m_a = \{20, 35, 55\}$ GeV -- passing the full event selection is shown in Figure~\ref{fig:HiggsZa:ATLAS_efficiencies_m34_1}, normalizing the signal yields to $\frac{1}{10}\sigma_{\rm SM}(pp \to h\to Z Z^{*} \to 4 \ell) = 0.69$ fb as done in~\cite{ATLAS:2021ldb}. We label this cross section as $\sigma_{0}$, which corresponds in our ALP signal model to ${\rm BR}(h\to Z a) \times {\rm BR}(a\to \mu\mu) = 2.11 \times 10^{-4} 
$.\footnote{Here we have used $\sigma( p p \to h) = 48.58$ pb, the N$^{3}$LO value for the gluon fusion production cross-section of the 125 GeV Higgs boson~\cite{Anastasiou:2016cez}, and BR$(Z \to \ell^+ \ell^-) = 0.06729$~\cite{ParticleDataGroup:2024cfk}  for the leptonic $Z$ branching fraction.}~We also display in Figure~\ref{fig:HiggsZa:ATLAS_efficiencies_m34_1} the dominant backgrounds for the search and the corresponding dark-photon $\sigma(p p\to h\to Z Z_d \to 4 \ell)$ signal predictions used in the ATLAS analysis~\cite{ATLAS:2021ldb} -- both obtained directly from Figure 9 of that analysis -- . 
%

\begin{figure}[h]
  \centering
  \includegraphics[width=0.67\textwidth]{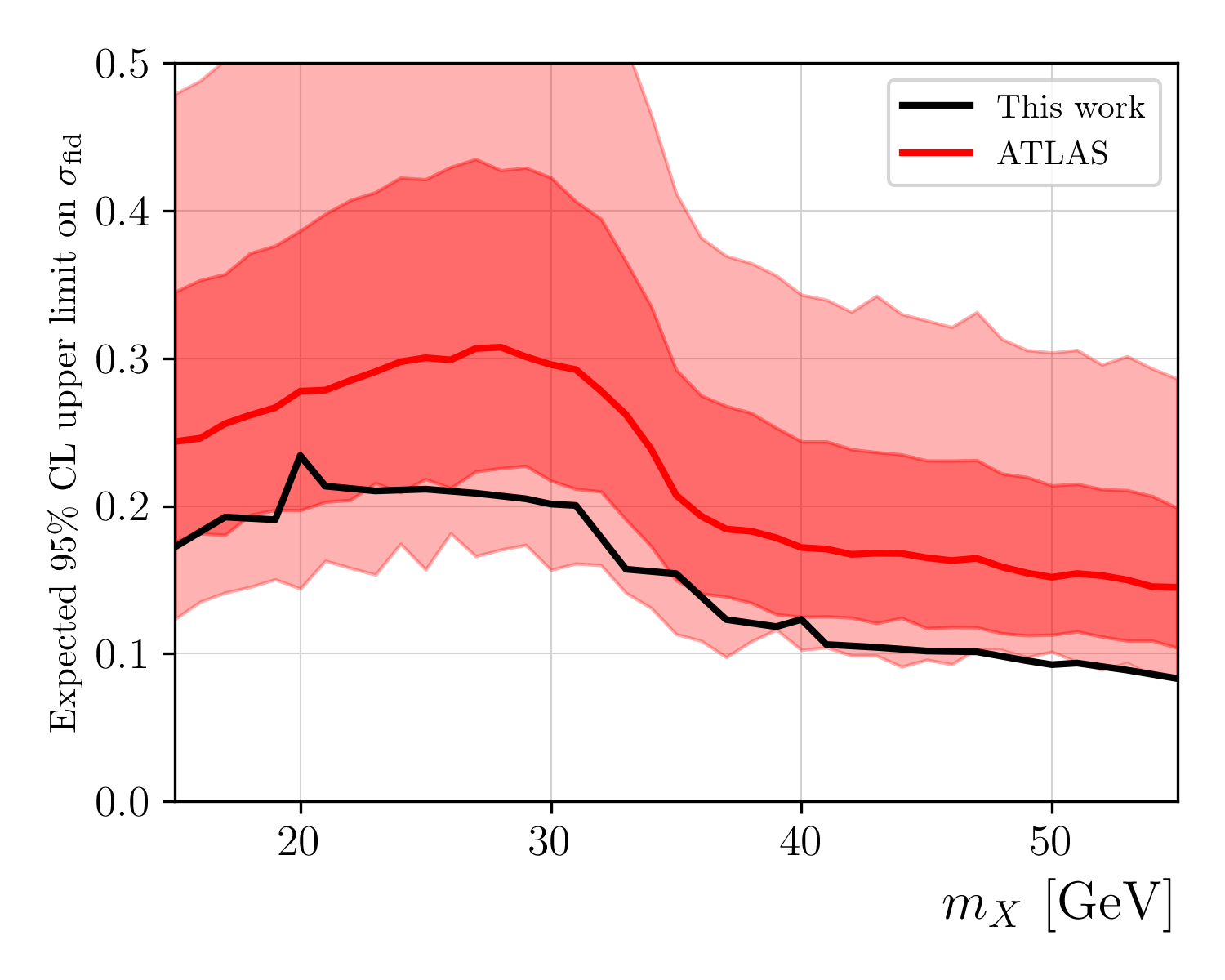}  

\vspace{-5mm}
  
  \caption{Upper (expected) limit at 95$\%$ C.L. on the fiducial cross-section $\sigma_{\rm fid} (gg \to h \to Z a \to \ell\ell\, \mu\mu)$, derived in this work (solid black line) as a function of $m_a = m_X$. We also include the corresponding expected limits on $\sigma_{\rm fid}$ from the ATLAS analysis~\cite{ATLAS:2021ldb} (solid red line) with their $\pm 1\sigma$ (dark red) and $\pm 2\sigma$ (light red) regions.}
\label{fig:HiggsZa:ATLAS_efficiencies_m34_2}
\end{figure}

We then use a binned likelihood analysis on this $m_{34}$ distribution to derive 95\% C.L. sensitivity limits on our signal cross section, in comparison with ATLAS. Specifically, we derive expected limits on the generator-level fiducial cross section $\sigma_{\rm fid} (gg \to h \to Z a \to \ell\ell\, \mu\mu)$, given by
\begin{equation}
\sigma_{\rm fid} = N_S^{95}/(\mathcal{L}\times\epsilon_c)    
\end{equation}
with $N_S^{95}$ our upper bound on the total number of signal events within the $m_{34}$ signal region, $\mathcal{L} = 139$ fb$^{-1}$ the integrated luminosity and $\epsilon_c$ the reconstruction efficiency. The value $N_S^{95}$ for each mass $m_a$ is obtained from a binned likelihood sensitivity analysis, with likelihood function $L(\mu)$ built as a product of bin Poisson probabilities~\cite{Cowan:2010js}, one for each bin of the $m_{34}$ distribution in Figure~\ref{fig:HiggsZa:ATLAS_efficiencies_m34_1}:
\begin{eqnarray}
L(\mu) = \prod_k \,e^{-(\mu \,s_k +\, b_k)}\, \frac{(\mu \,s_k + b_k)^{n_k}}{n_k !}   
\label{likelihood_NS}
\end{eqnarray}
where $n_k$ is the number of observed events in each bin, and $b_k$, $s_k$ are respectively the number of SM background and signal -- for a signal cross section $\sigma_0$ -- events in each bin. The signal strength parameter $\mu$ is given by $\mu = N_S^{95}/(\sum_k s_k)$, and we make $n_k = b_k$ to determine the expected exclusion sensitivity in the absence of a signal. We define our test statistics $Q_{\mu}$ as
\begin{eqnarray}
Q_{\mu} = -2\, \mathrm{Log} \left[\frac{L(\mu)}{L(\hat{\mu})} \right]
\label{likelihood_1}
\end{eqnarray}
with $\hat{\mu}$ being the value of $\mu$ which maximizes $L(\mu)$. We note that our analysis does not include the effect of background systematic uncertainties. The value of $\mu$ that can be excluded at 95\% C.L. -- denoted by $\mu^{max}$ -- is given by
$Q_{\mu} = 3.84$. Our resulting expected 95\% C.L. exclusion limits on $\sigma_{\rm fid} (gg \to h \to Z a \to \ell\ell\, \mu\mu)$
are shown in Figure~\ref{fig:HiggsZa:ATLAS_efficiencies_m34_2}, compared to those from the ATLAS analysis (Fig.~16 of~\cite{ATLAS:2021ldb}). Our limits are somewhat stronger than those of ATLAS, 
yet they display the same trend with $m_a$ and are within the ATLAS 2$\sigma$ band for (almost) the entire $m_X$ range covered in the analysis.

\begin{figure}[h]
  \centering
  \includegraphics[width=0.7\textwidth]{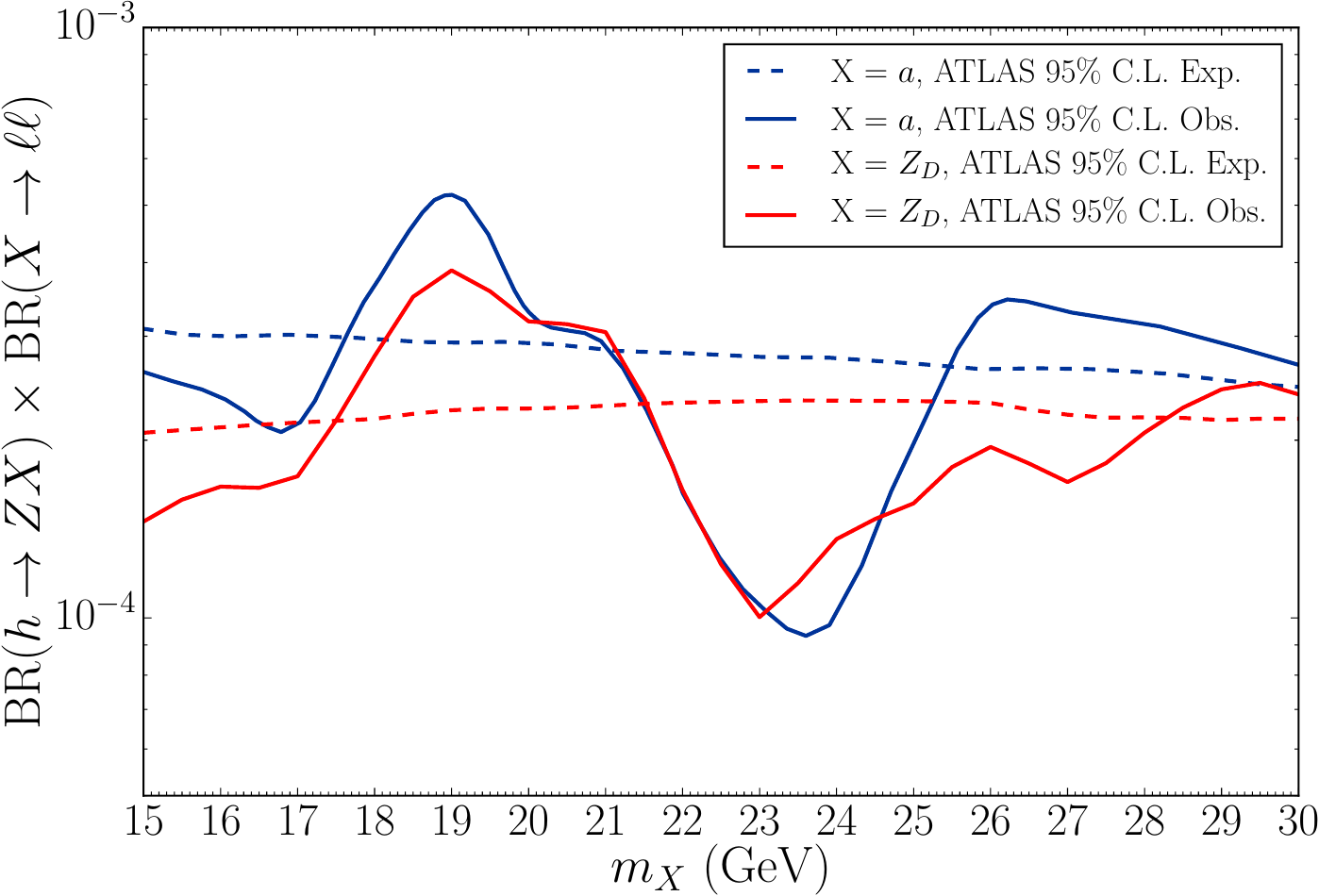}  
  \caption{ATLAS expected (dashed) and observed (solid) 95\% C.L. upper limit on ${\rm BR}(h\to Z a) \times {\rm BR}(a\to \mu\mu)$ (blue) as a function of $m_a = m_X$. We also show for comparison the corresponding expected and observed limits on ${\rm BR}(h\to Z Z_D) \times {\rm BR}(Z_D\to \ell\ell)$ for the dark photon model used by ATLAS (red).}
\label{fig:HiggsZa:ATLAS_limits_BR}
\end{figure}

Finally,  the expected and observed ATLAS cross section limits from~\cite{ATLAS:2021ldb} -- Fig.~17 (b) of the ATLAS analysis -- can be simply cast as (expected and observed) 95\% C.L. upper limits on ${\rm BR}(h\to Z a) \times {\rm BR}(a\to \mu\mu)$ as a function of the pseudoscalar mass $m_a$ in our ALP model, for $m_a \in [15, 30]$ GeV. These limits are shown in Figure~\ref{fig:HiggsZa:ATLAS_limits_BR}, together with the corresponding limits on ${\rm BR}(h\to Z Z_D) \times {\rm BR}(Z_D\to \ell\ell)$ extracted from Fig.~17 (a) of the ATLAS analysis for comparison (which shows that the acceptances for both models are indeed fairly similar). The ALP limits from Figure~\ref{fig:HiggsZa:ATLAS_limits_BR} will be used later in Section~\ref{sec:experiment} when we discuss the interplay between several $h\to Z a$ search channels.

\section{\texorpdfstring{\textbf{CMS $\bm{h \to a a \to \mu \mu \tau \tau}$ search validation (same final state)}}{CMS h → aa → μμττ search validation (same final state)}}
\label{sec:AppII}

As outlined in Section~\ref{sec:1}, searches for exotic Higgs decays in the $\mu \mu \tau \tau$ final state have been performed by both ATLAS and CMS collaborations~\cite{ATLAS:2015unc,CMS:2018qvj,CMS:2020ffa}. Nevertheless, the signal selection exclusively targets the $h \to a a$ decay mode, and has very limited sensitivity to the $h\to Z a$ exotic decay (see e.g.~\cite{Brooijmans:2020yij}) which is the focus of this work. Yet, in this section, we aim at reproducing these experimental results, as a means to calibrate our subsequent analysis in Section~\ref{sec:experiment}. We focus on the CMS $h \to a a \to 2\mu2\tau$ analysis~\cite{CMS:2018qvj} performed for 13 TeV LHC data with an integrated luminosity of 35.9 fb$^{-1}$, as it provides enough public information to attempt reproducing their results.

The CMS search starts by considering only muons with $p_{T} > 5$ GeV and $|\eta| < 2.4$, electrons with $p_{T} > 7$ GeV and $|\eta| < 2.5$, and tau-tagged hadronic jets with $p_{T} > 18.5$ GeV and $|\eta| < 2.3$. The analysis uses three different online muon triggers to select events:
\begin{itemize}
    \item Single-muon trigger: leading muon with $p_{T} > 24$ GeV.
    \item Double-muon trigger: leading muon with $p_{T} > 17$ GeV, and sub-leading muon with $p_{T} > 8$ GeV.
    \item Triple-muon trigger: leading muon with $p_{T} > 12$ GeV, and sub-leading muon with $p_{T} > 10$ GeV.
\end{itemize}

At the offline level, the analysis imposes a leading muon $p_{T} > 18$ GeV (or $p_{T} > 25$ GeV if the event only passes the single-muon trigger) and a subleading muon $p_{T} > 9$ GeV (or $p_{T} > 11$ GeV if the event only passes the triple-muon trigger). Each selected event must include at least an opposite sign (OS) pair of muons and an OS 
pair of tau-candidates, which can respectively correspond to $e \mu$, $e \tau_{h}$, $\mu \tau_h$, or $\tau_{h} \tau_{h}$ (with $\tau_{h}$ being a tau-tagged hadronic jet) final states. The highest-$p_{T}$ muon and next-to-highest $p_{T}$ OS muon are considered to have originated from one of the intermediate $a$ resonances. If there is an electron and/or a third muon, these are considered a decay product of the $\tau$-lepton(s).
All electrons and muons in the event must satisfy the $\Delta R > 0.3$ isolation requirement, or $\Delta R > 0.4$ in case there is a $\tau_{h}$ in the event ($\tau_{h}$ candidates are built from jets with distance parameter of $\Delta R = 0.4$). If more isolated electrons or muons than the ones needed to form the di-muon and di-tau candidate pairs, then the event is discarded from the analysis. 

The above represents the baseline final state selection of the CMS analysis. In addition, the CMS signal selection requires the di-muon invariant mass to be within the range 14 GeV $< m_{\mu\mu} <$ 64 GeV, together with a $\tau$-pair visible mass $m_{\tau\tau}^{\rm vis}$ -- this is the invariant mass of the visible objects comprising the pair of tau-candidates, as discussed above -- smaller than the di-muon invariant mass. 
Finally, to further suppress the background the CMS analysis requires the visible invariant mass $m_{\rm vis}$ of the four particles:
\begin{itemize}
    \item $m_{\rm vis} < 110$ GeV for the $\mu \mu + e \mu$ final state.
    \item $m_{\rm vis} < 120$ GeV for the $\mu \mu + e \tau_{h}$ and $\mu \mu + \mu \tau_{h}$ final states.
    \item $m_{\rm vis} < 130$ GeV for the $\mu \mu + \tau_{h} \tau_{h}$ final state.
\end{itemize}
This last selection cut is justified by the fact that the visible invariant mass is expected to shift from the $m_h = 125$ GeV peak depending on the number of neutrinos from the $\tau$-lepton decays. Additionally, a b-jet veto with $p_{T} > 20$ GeV is applied to suppress backgrounds with b-tagged jets, such as $t\bar{t}$.

\begin{figure}[h]
    \centering

    \includegraphics[width=0.65\textwidth]{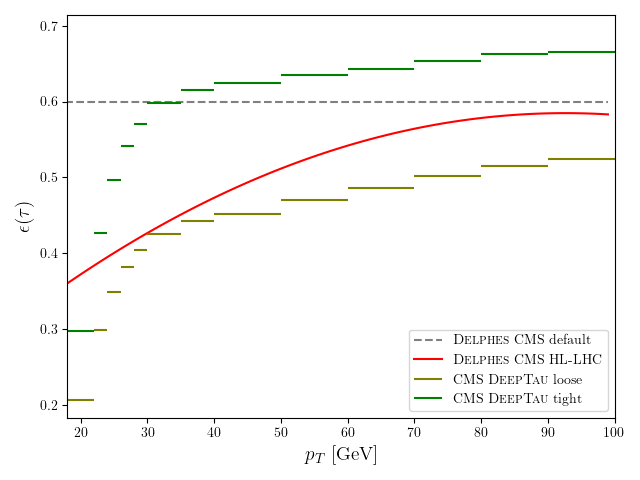}

\vspace{-4mm}
    
    \caption{$\tau$-tagging efficiency $\et$ of the \texttt{TauTagging} module in {\sc Delphes}, for different configurations used in this work. We show the ``loose'' (olive green) and ``tight'' (dark green) working points from the CMS {\sc DeepTau} algorithm in Ref.~\cite{CMS:2022prd}. We also show the {\sc Delphes} CMS HL-LHC parametrization (red) used in Section~\ref{sec:models}, and the default {\sc Delphes} parametrization (dashed-grey) for comparison.}
    \label{fig:taueff}

\end{figure}

\begin{figure}[t!]
    \centering
    \includegraphics[width=.42\linewidth]{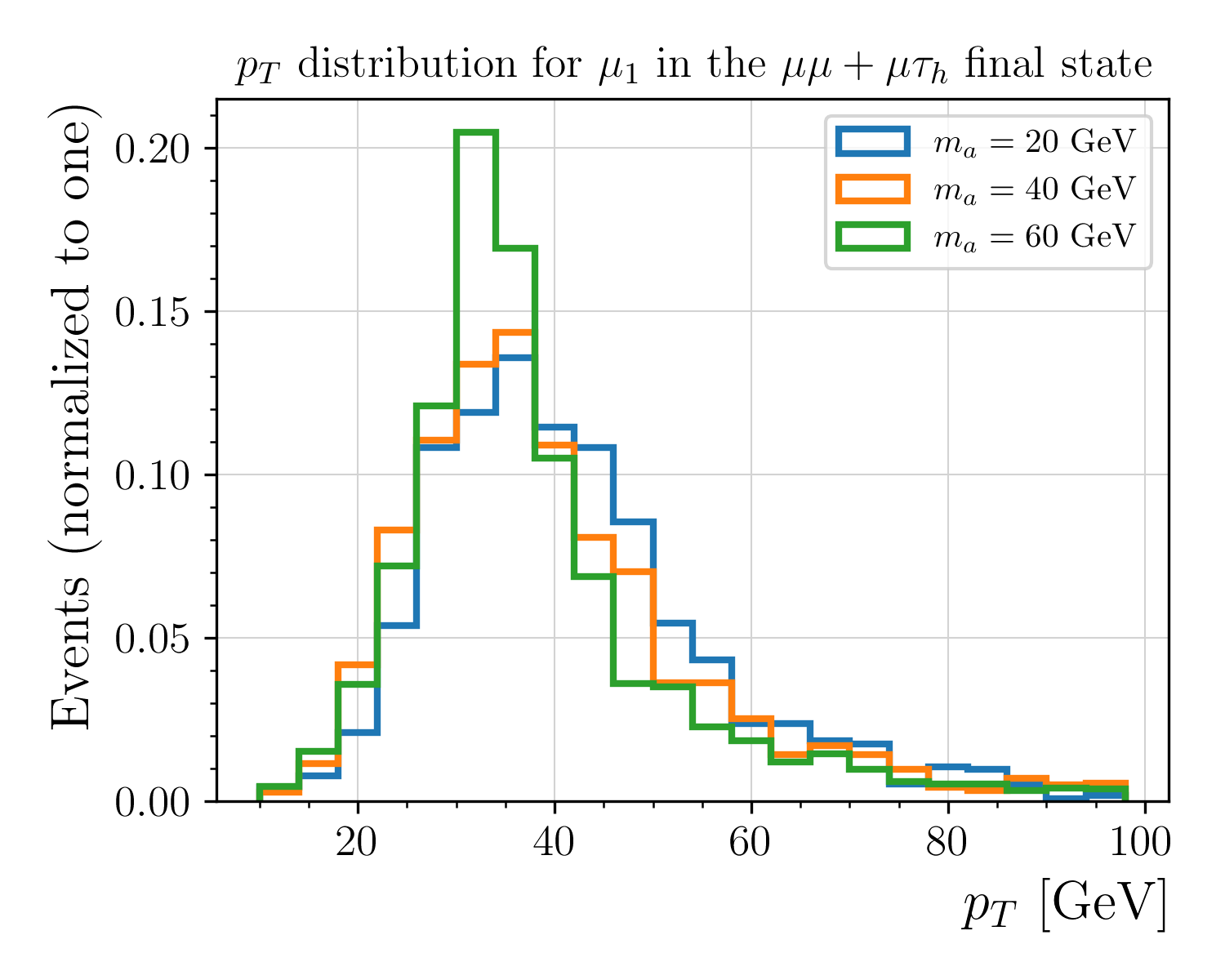}
    \includegraphics[width=.42\linewidth]{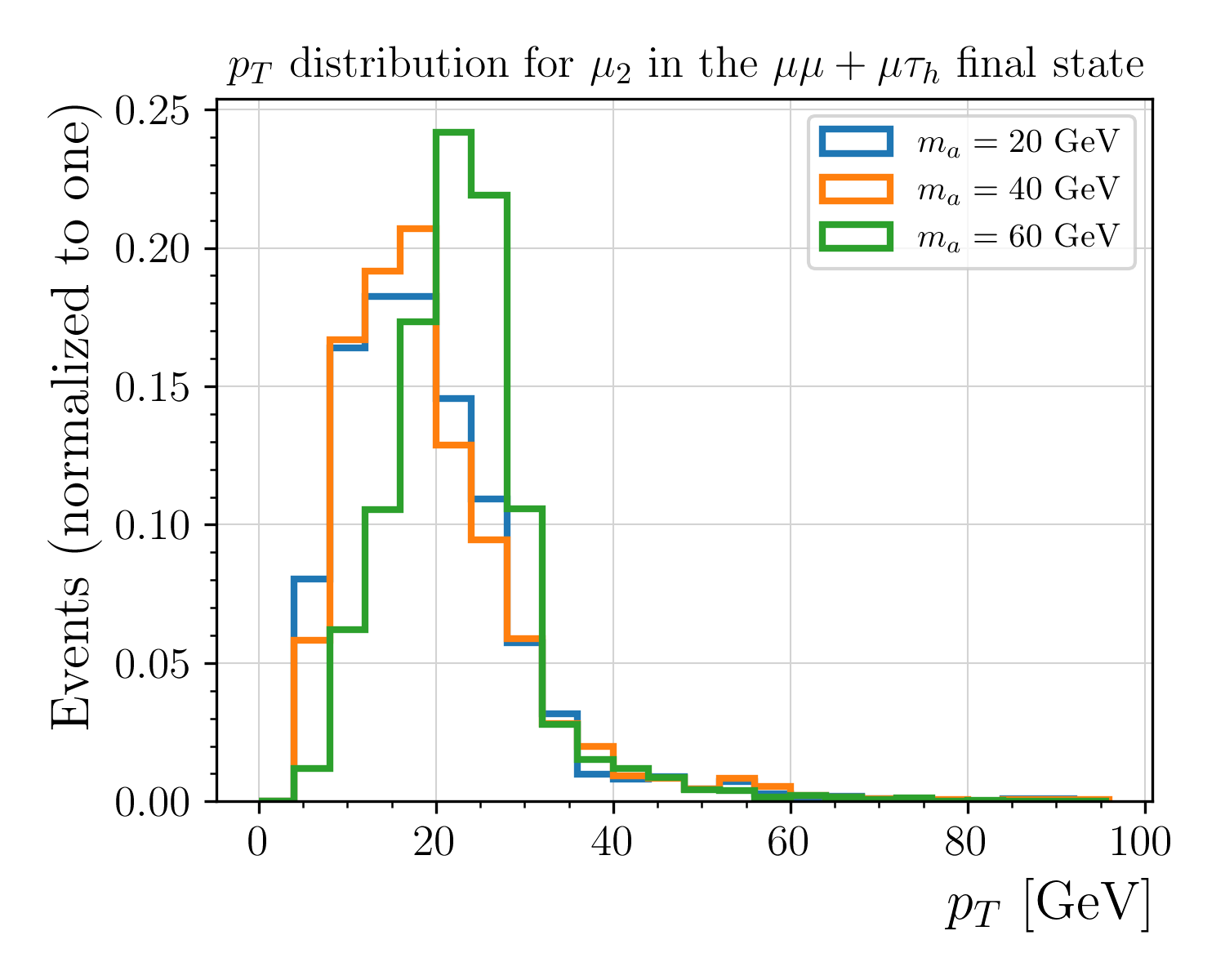}
    \includegraphics[width=.42\linewidth]{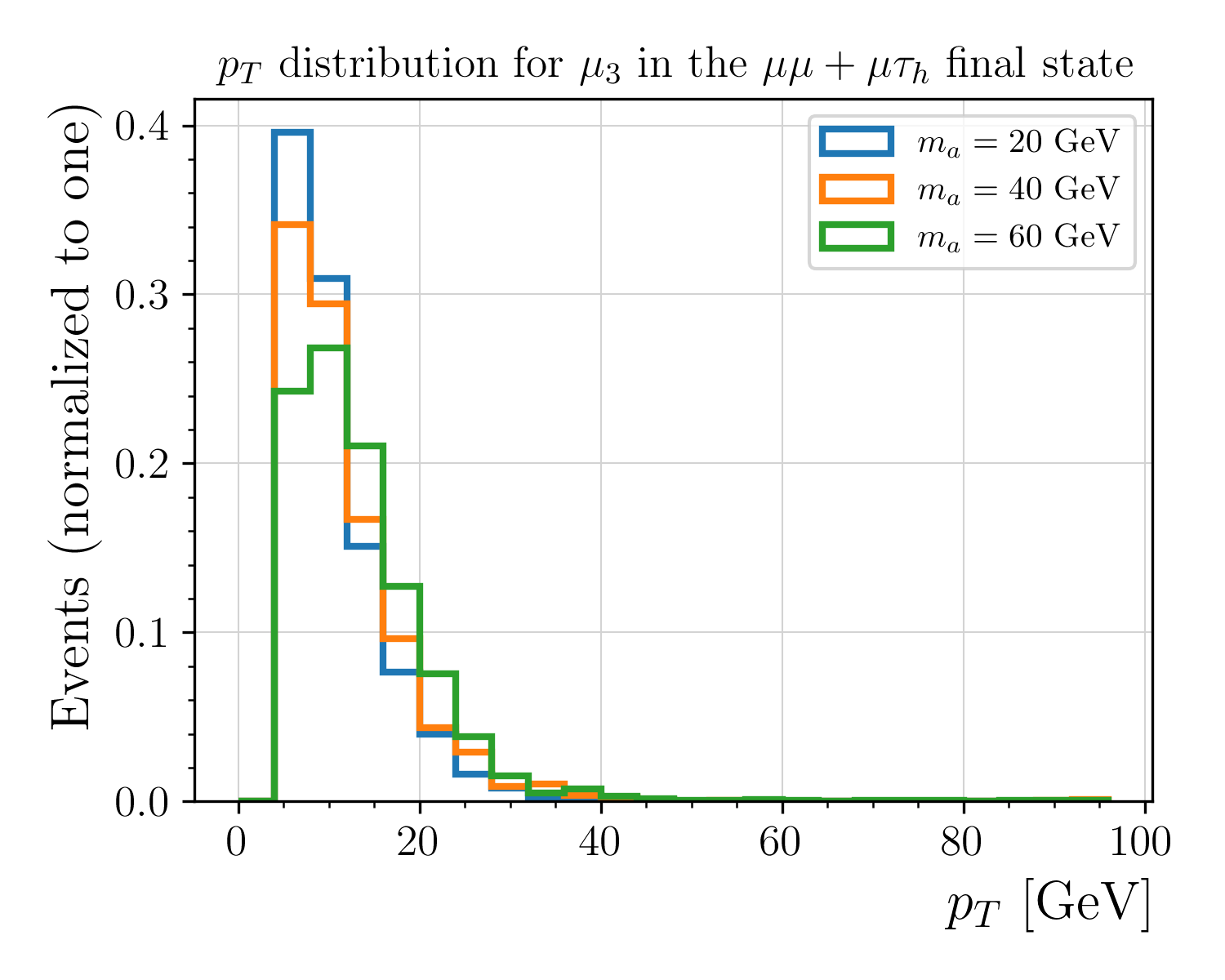}
    \includegraphics[width=.42\linewidth]{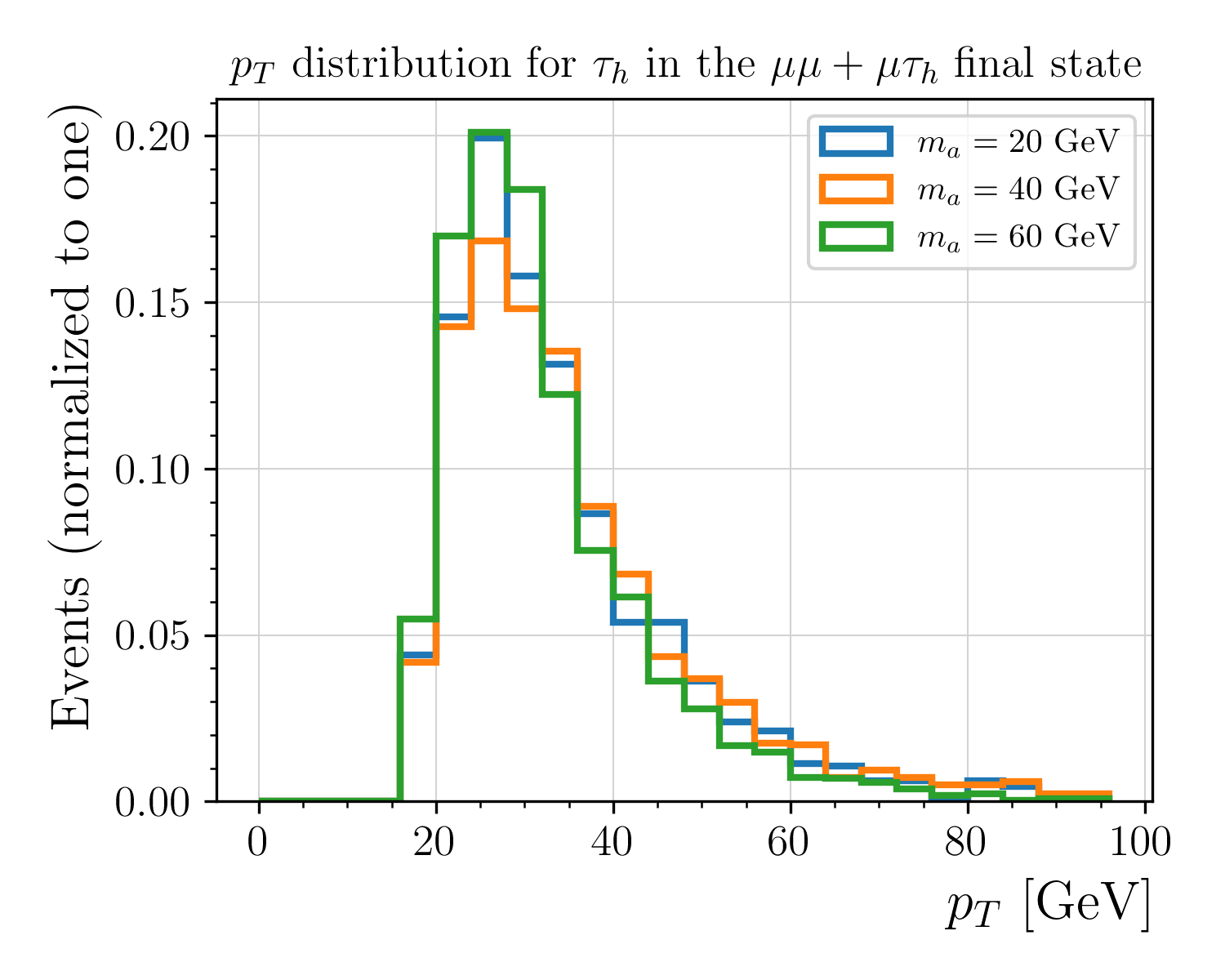}

    \vspace{-4mm}
    
    \caption{$p_{T}$ distribution in the $\mu\mu + \mu \tau_{h}$ final state, for the 
    leading-$p_{T}$ muon $\mu_1$ (top-left), the subleading-$p_{T}$ muon $\mu_2$ (top-right), the third muon $\mu_3$ (bottom-left) and the hadronically decaying tau $\tau_h$ (bottom-right)
    after the baseline event selection of CMS~\cite{CMS:2018qvj} (and before further signal selection cuts on kinematical variables), for $m_a = 20$ GeV (blue), $m_a = 40$ GeV (orange) and $m_a = 60$ GeV (green).}
    \label{fig:pt-mumumutau}
\end{figure}

For our signal event generation, we use the same UFO model as in Section~\ref{sec:AppI}, together with {\sc MadGraph5\_aMC@NLO} + {\sc Pythia8} + {\sc Delphes}. As opposed to the CMS analysis~\cite{CMS:2018qvj}, we generate only $p p \to h \to a a \to 2 \mu \, 2 \tau$ signal events -- and do not consider the signal contribution from the $p p \to h \to a a \to 4 \tau$ channel (with at least two tau-leptons decaying leptonically), since this channel will be irrelevant for our newly proposed search (see the discussion in Section~\ref{sec:experiment}) -- .\footnote{Besides, generating a statistically meaningful sample of signal events for the $p p \to h \to a a \to 4 \tau$ channel is a very resource-intensive process, as only a tiny fraction of the generated events satisfies the CMS analysis event selection.} For our detector simulation, we use a modified version of the standard {\sc Delphes} CMS card, with the transverse-momentum thresholds for electrons and muons respectively set to $p_{T} > 7$ GeV and $p_{T} > 5$ GeV, to match the CMS analysis selection. Furthermore, electrons are reconstructed for $|\eta| \leq 2.5$, and identified with an isolation cone of $\code{DeltaRMax} = 0.3$ in the {\sc Delphes} card $\code{ElectronIsolation}$ module.\footnote{We would like to thank Cecile Caillol for clarifying correspondence about this point.} Muons are reconstructed for $|\eta| \leq 2.7$, and we set 
$\code{PTRatioMax} = 0.2$ for the relative isolation and $\code{DeltaRMax} = 0.4$ in the $\code{MuonIsolation}$ module, following~\cite{CMS:2018qvj}. The muon efficiency in the $\code{MuonEfficiency}$ module is also set to 0.98. In addition, the minimum $p_{T}$ of reconstructed jets is set to $16$ GeV -- slightly below the $p_{T} > 18.5$ GeV threshold for tau-tagged jets in the CMS event selection --. For the reconstruction of tau-tagged jets, we set $\code{DeltaR}=0.4$ and $\code{TauEtaMax}=2.3$ in the $\code{TauTagging}$ module, and implement a parametrization of the tau-tagging efficiency $\epsilon(\tau)$ following the ``loose'' and ``tight'' working points from the CMS {\sc DeepTau} algorithm~\cite{CMS:2022prd} (binned in $p_T$)\footnote{We note that the $p_T$ threshold for tau reconstruction in~\cite{CMS:2022prd} corresponds to 20 GeV. We extend this threshold down to $p_{T} > 18.5$ GeV, to match the minimum threshold of the CMS $p p \to h \to a a \to 2 \mu \, 2 \tau$ analysis~\cite{CMS:2018qvj}.} as shown in Figure~\ref{fig:taueff}. In addition, we compare in the same figure the default and the CMS HL-LHC parametrization~\cite{ATLAS:2022hsp}, the latter being encompassed by the ``loose'' and ``tight'' configurations. Specifically, we adopt in our analysis the ``tight'' CMS working point for $\epsilon(\tau)$. We also implement the corresponding ``tight'' working point fake-tau efficiency \ejt from~\cite{CMS:2022prd}, which is below 0.003 for the entire $p_T$ range.

\begin{figure}[h!]
    \centering
    \includegraphics[width=.65\linewidth]{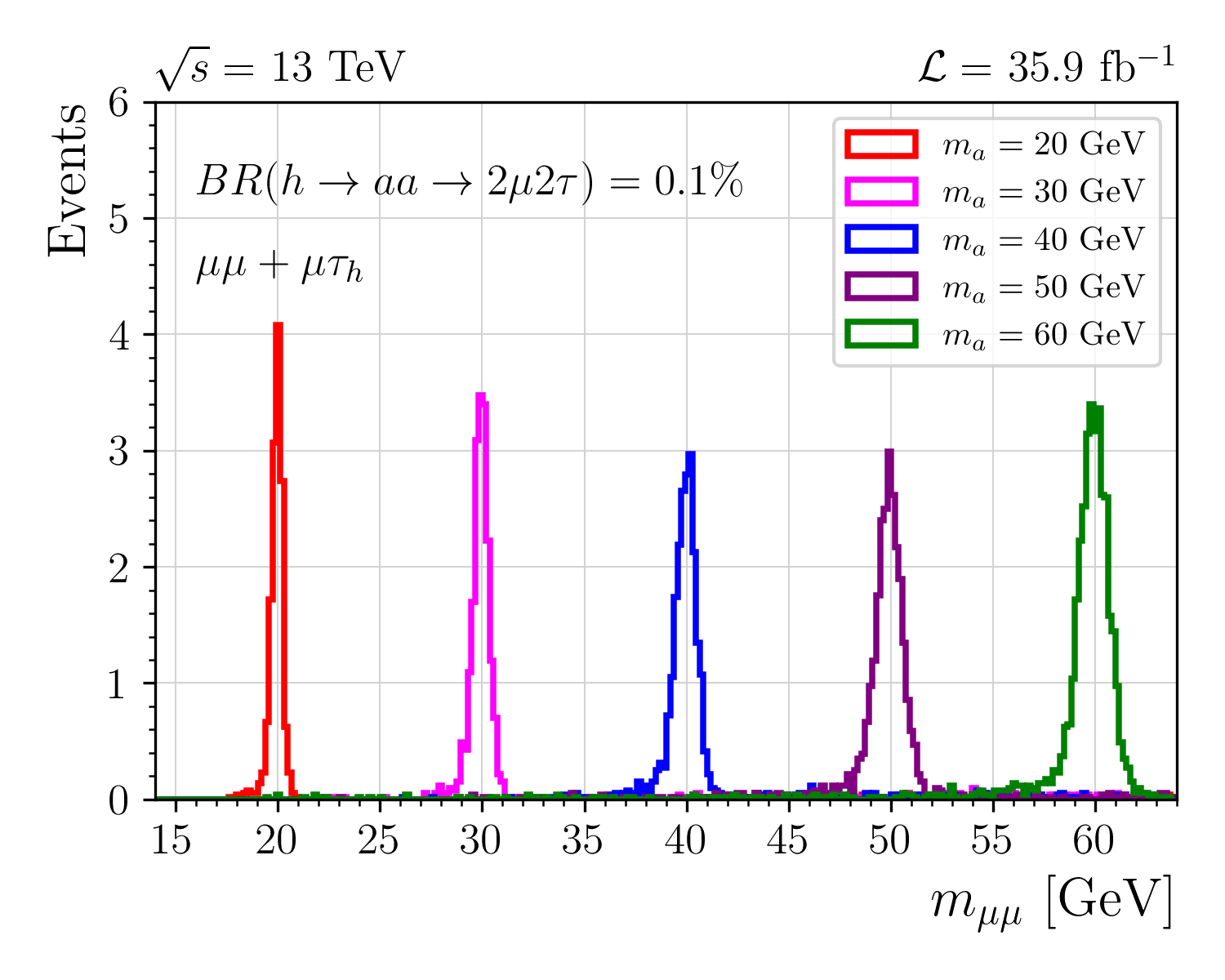}

\vspace{-4mm}

    \caption{Signal yields for the $h \to aa \to 2\mu2\tau$ process after event selection, as a function of the invariant mass of the di-muon system $m_{\mu\mu}$, for $m_a = 20$ GeV (red), $m_a = 30$ GeV (pink), $m_a = 40$ GeV (blue), $m_a = 50$ GeV (purple), $m_a = 60$ GeV (green), for the  $\mu \mu + \mu \tau_{h}$ final state.}
    \label{fig:signal-comparison}
\end{figure}

In Figure~\ref{fig:pt-mumumutau} we show, for the $\mu\mu + \mu \tau_{h}$ final state, the (reco-level) $p_{T}$ distribution of each particle in the final state -- the leading-$p_{T}$ muon $\mu_1$, the subleading-$p_{T}$ muon $\mu_2$, the third muon $\mu_3$ (which is assumed to come from a $\tau$ decay) and the hadronically decaying tau $\tau_h$ -- for $m_a = 20$ GeV, $40$ GeV and $60$ GeV --, after the baseline final state selection of the CMS analysis (but without further signal selection cuts). Then, in Figure~\ref{fig:signal-comparison} we show the surviving signal events after signal selection cuts -- for $\sigma( p p \to h) = 48.58$ pb~\cite{Anastasiou:2016cez} and a branching fraction BR($h\to a a \to 2 \mu \, 2 \tau$) $= 0.1\%$ -- as a function of the $m_{\mu\mu} $ invariant mass, for different values of $m_a$. Figure~\ref{fig:signal-comparison} displays an excellent agreement with Fig. 1 (left) of~\cite{CMS:2018qvj}.
As a final check of our ability to reproduce the results from~\cite{CMS:2018qvj}, we show in Table~\ref{tab:signal-yield-tota} the 
number of $p p \to h \to a a \to \mu\mu \tau\tau$ events after signal selection cuts for each of the four final states of the analysis, $\mu \mu + e \mu$, $\mu \mu + e \tau_{h}$, $\mu \mu + \mu \tau_{h}$, and $\mu \mu + \tau_{h} \tau_{h}$, for $m_a = 20$, $40$, $60$ GeV -- for $\text{BR}(h \to aa \to 2\mu2\tau) = 0.01 \%$ -- , together with the corresponding numbers in the CMS analysis. Our agreement with the CMS analysis is very good for $m_a = 60$ GeV across all final states, and for the other masses in the $\mu\mu + \mu \tau_{h}$ and $\mu\mu + e\mu $ final states -- better than 20\% in all these cases -- , while in the remaining cases the difference is still within a factor $2$.

\begin{table}[h!]
\centering
\begin{tabular}{cc|c|c|c|c}
\hline
     & & $\mu \mu + e \mu$ & $\mu \mu + e \tau_{h}$ & $\mu \mu + \mu \tau_{h}$ & $\mu \mu + \tau_{h} \tau_{h}$ \\ 
\cline{1-6}
\multirow{2}{*}{$m_a$ = 20 GeV}  &  This work~& 0.44 & 0.35 & 0.56 & 0.15 \\ 
                                &  CMS (13 TeV, 35.9 fb$^{-1}$) & 0.39 & 0.25 & 0.47 & 0.10 \\ 
                                \cline{1-6}
\multirow{2}{*}{$m_a$ = 40 GeV } & This work & 0.61 & 0.57 & 0.86 & 0.29 \\  
                                & CMS (13 TeV, 35.9 fb$^{-1}$) & 0.57 & 0.28 & 0.68 & 0.14 \\ \cline{1-6}
\multirow{2}{*}{$m_a$ = 60 GeV } &  This work & 0.86 & 1.01 & 1.34 & 0.57 \\  
                               &  CMS (13 TeV, 35.9 fb$^{-1}$) & 0.94 & 0.85 & 1.18 & 0.52 \\ 
\cline{1-6}
\end{tabular}
\caption{Signal yields for the $h \to aa \to 2\mu2\tau$ process considering $\text{BR}(h \to aa \to 2\mu2\tau) = 0.01 \%$, for the four final states, compared to CMS~\cite{CMS:2018qvj} expected values for 13 TeV and 35.9 fb$^{-1}$.}
\label{tab:signal-yield-tota}
\end{table}

Having validated the CMS search for $h \to a a \to \mu\mu\tau\tau$ provides insights into the kinematic properties and reconstruction challenges of the final state we will later explore. While this search targets a different intermediate state than our proposed analysis, the fact that it shares the same final state at reco-level allows us to understand the detector response, particularly the efficiency and limitations of tau-lepton reconstruction. This validation also enables us to precisely calibrate our detector simulation setup in {\sc Delphes}, ensuring that our modeling of object reconstruction and event selection criteria will closely reflect current experimental conditions.


\section{\texorpdfstring{\textbf{New LHC searches in $\bm{h \to Z a \to \ell \ell \tau \tau}$}}{New LHC search in h → Z a → llττ}}
\label{sec:experiment}

Having studied in the previous sections the existing ATLAS and CMS searches for exotic Higgs decays targeting either the same intermediate state $h\to Z a$ or the same final state $h\to \mu\mu \tau\tau$, we can turn now to perform our analysis for the exotic Higgs decay $h \to Z a \to \ell\ell \tau\tau$ (with $\ell = e,\,\mu$) -- considering both hadronic and leptonic decays of the $\tau$-leptons -- , as a means to demonstrate the power of the $a \to \tau\tau$ decay to probe BSM scenarios.  We concentrate in the following on the $\mu\mu \tau\tau$ final state due to its cleanliness, yet stress that including the $e e \tau\tau$ final state in the analysis is also possible and would increase the sensitivity of our proposed search.

To generate our $p p \to h \to Z a \to \mu\mu \tau\tau$ signal events, we use the UFO model already adopted in Sections~\ref{sec:AppI} and~\ref{sec:AppII}, and the {\sc MadGraph5\_aMC@NLO} + {\sc Pythia8} + {\sc Delphes} event-generation pipeline, this time considering an LHC center-of-mass energy of $\sqrt{s}=14$ TeV. For the sake of simplicity, the LHC production of the Higgs boson $h$ is simulated including only gluon-fusion.
For our detector simulation, we use here the same {\sc Delphes} card as in the validation performed in Section~\ref{sec:AppII} (as we are dealing with the same reco-level final state), only changing the $\tau$-tagging efficiency $\epsilon(\tau)$, for which we now use the Phase-2 CMS parametrization given in~\cite{ATLAS:2022hsp} for HL-LHC. This $\epsilon(\tau)$ is shown explicitly in Figure~\ref{fig:taueff} (red line).

As SM background processes, we consider the irreducible background $p p \to \mu \mu \tau \tau$, which arises from $p p \to Z Z^{(*)} $ and from SM Higgs production through gluon-fusion $p p \to h \to  Z Z^{*}$ (with $Z Z^{(*)} \to \mu \mu \tau \tau$), as well as from SM Higgs associated production $p p \to Z h$, with $Z \to \mu \mu$ and $h\to \tau\tau$. We simulate the three contributions separately, and for the former (which constitutes the dominant background) we perform the matching in 
up to one extra hard jet using the MLM prescription~\cite{Mangano:2002ea,Alwall:2007fs}. We also consider reducible SM backgrounds, $p p \to t\bar{t}Z$ (with $Z \to \mu \mu$) and $p p \to \mu \mu \tau \tau + X$ (with $X=2 \ell,\, 2 \nu$),\footnote{We have not considered other reducible SM backgrounds such as $Z/W$ + jets, $ZW$ and $t\bar{t}$, whose estimate would require an accurate treatment of fake-muons and/or fake-taus which is beyond our capabilities (and likewise for instrumental background). We nevertheless expect these SM backgrounds would be included in a full-fledged experimental analysis, and their importance quantified.} yet we find that these reducible backgrounds combined yield less than 0.1 expected events in the final signal region of our analysis, and can then be safely disregarded.

\begin{table}[]
    \centering
    \begin{tabular}{c|c|c|c|c }
         \hline 
         Selection & \makecell{Signal \\ ($m_a = 20$ GeV)} & \makecell{$pp \to 2\mu 2\tau$ \\ ($0, 1$ $j$ matched)}  & $gg \to h \to 2\mu 2\tau$ & \makecell{$pp \to Z h$ \\ ($h \to \tau\tau$)} \\ \hline 
         Inclusive $\sigma$ (fb) & $0.74$ & $22.5$ & $9.57$ & $0.83$ \\
         \hline 
         Online $\mu$-trigger & $0.62$ & $20.8$ & $2.98$ & $0.71$ \\
         \hline 
         Offline $\mu$-trigger & $0.41$ & $17.4$ & $1.63$ & $0.48$ \\ 
         \hline 
         \makecell{OS $\mu\mu$ pair \\ + $\tau$-candidate pair \\ + $\Delta R$ isolation} &  $0.0037$ & $2.72$ & $0.052$ & $0.083$ \\
         \hline 
         $m_{\rm vis}$ selection & $0.0018$ & $0.087$ & $0.043$ & $1.43 \times 10^{-4}$ \\
         \hline 
         $m_{\tau \tau}^{\rm{vis}} < m_{\mu \mu}$ & $0.0018$ & $0.036$ & $0.018$ & $1.03 \times 10^{-4}$ \\
         \hline 
         $m_{\mu \mu}$ selection & $0.0017$ & $0.015$ & $0.0075$ & $8.01 \times 10^{-5}$ \\
         \hline 
         $b$-jet veto & $0.0016$ & $0.015$ & $0.0071$ & $7.84 \times 10^{-5}$ \\
         \hline
    \end{tabular}
    
    \caption{Analysis cutflow of the cross section (in fb) for the $p p \to h \to Z a \to \mu\mu\, \tau\tau$ signal -- for  $m_{a} = 20$ GeV and BR$(h \to Z a) \times$ BR$(a \to \tau\tau) = 4 \times 10^{-4}$ -- and dominant SM backgrounds: $p p \to Z Z^{(*)} \to \mu \mu \tau \tau$ (matched up to one extra hadronic jet), gluon-fusion SM Higgs production $p p \to h \to \mu \mu \tau \tau$, and associated SM Higgs production $p p \to Z h$ with $Z\to \mu\mu$ and $h\to \tau\tau$. All other SM backgrounds are negligible (see text for details).}
    \label{tab:cutflow1}
\end{table}

For the event selection criteria in our analysis, we have applied a combination of requirements from the validated analysis from Sections \ref{sec:AppI} and~\ref{sec:AppII}, with a large portion of the selection cuts inspired by the $h \to a a \to \mu\mu\,\tau\tau$ CMS search~\cite{CMS:2018qvj} (recall Section~\ref{sec:AppII}), as it shares the same final states at {\sc Delphes} level. 

Specifically, we select muons with $p_{T} > 5$ GeV and $|\eta| < 2.4$, electrons with $p_{T} > 7$ GeV and $|\eta| < 2.5$, and $\tau$-tagged jets with $p_{T} > 18.5$ GeV and $|\eta| < 2.3$. We stress that the latter $p_T$-threshold for hadronically decaying $\tau$-leptons ($\tau_h$), optimized for CMS Run 2, is the main responsible for the loss of sensitivity at low values of $m_a$ in our analysis, and a potential reduction in this threshold by ATLAS/CMS at the HL-LHC (e.g. down to $p_{T} > 16$ GeV) could significantly extend the coverage of our proposed search in final states including $\tau_h$.
As initial selection, we have adopted the same online triggers (single-muon, double-muon and triple-muon)  
and the same baseline final-state selection criteria -- including the offline trigger $p_T$-thresholds for muons, the requirement of an OS pair of muons and a pair of tau-candidates (yielding $e \mu$, $e \tau_{h}$, $\mu \tau_h$ or $\tau_{h} \tau_{h}$ final states), and the $\Delta R$ isolation criteria for each reconstructed object -- used in the CMS analysis in Section~\ref{sec:AppII}. These represent the baseline final-state selection criteria of our analysis. Subsequently, the di-muon invariant mass $m_{\mu\mu}$ is built out of the OS muon pair whose invariant mass is closest to the $Z$-boson mass.
We require $m_{\tau\tau}^{\rm vis} < m_{\mu\mu}$ (with $m_{\tau\tau}^{\rm vis}$ the visible mass of the pair of tau-candidates) and adopt the same requirement on the visible invariant mass $m_{\rm vis}$ of the four particles as in Section~\ref{sec:AppII}: $m_{\rm vis} < 110$ GeV ($\mu \mu + e \mu$ final state), $m_{\rm vis} < 120$ GeV ($\mu \mu + e \tau_{h}$ and $\mu \mu + \mu \tau_{h}$ final states), $m_{\rm vis} < 130$ GeV ($\mu \mu + \tau_{h} \tau_{h}$ final state). In addition, we demand the di-muon pair to be in the mass window 50 GeV $ < m_{\mu\mu} < 106$ GeV (as in the ATLAS search~\cite{ATLAS:2021ldb} validated in Section~\ref{sec:AppI}). This window allows not only to capture off-shell effects in $h\to Z a$ for pseudoscalar masses $m_a > 20$ GeV, but also ensures a sufficient number of SM background events in all $\tau\tau$ final states for our sensitivity estimates to be statistically robust. We have verified that a narrower mass window, e.g. $m_{\mu\mu} \in [m_Z - 10\,{\rm GeV},\, m_Z + 10\,{\rm GeV}]$ may result in a sensitivity increase only for $m_a \leq 20$ GeV, while it yields too few SM background events, potentially making our binned likelihood analysis unreliable.
Finally, we veto events with $b$-tagged jets with $p_{T}^j > 20$ GeV, to suppress SM backgrounds including top quarks (e.g. $t\bar{t}$ or $t\bar{t} Z$). 

\begin{figure}[t]
    \centering   
    \includegraphics[width=0.495\linewidth]{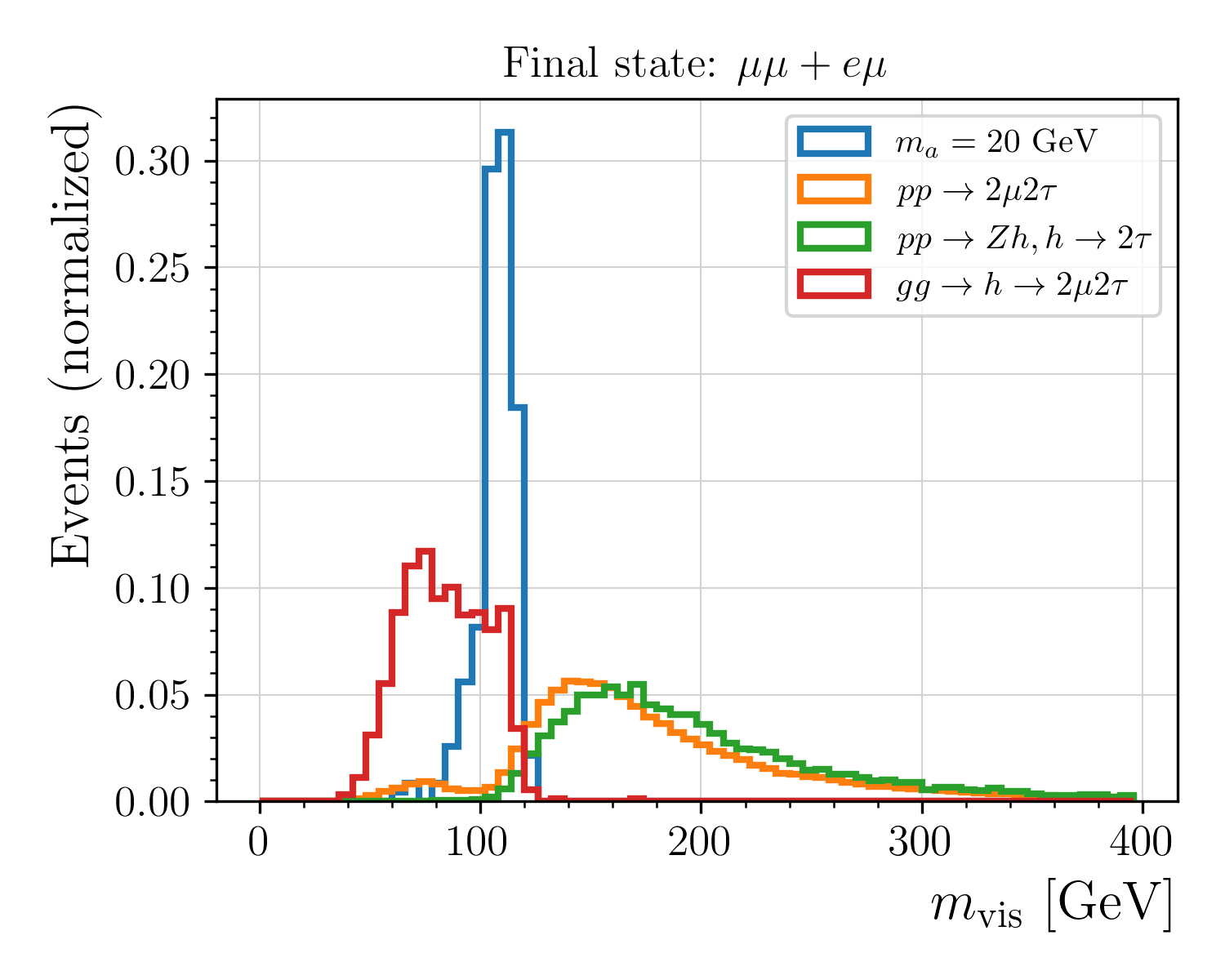}
    \includegraphics[width=0.495\linewidth]{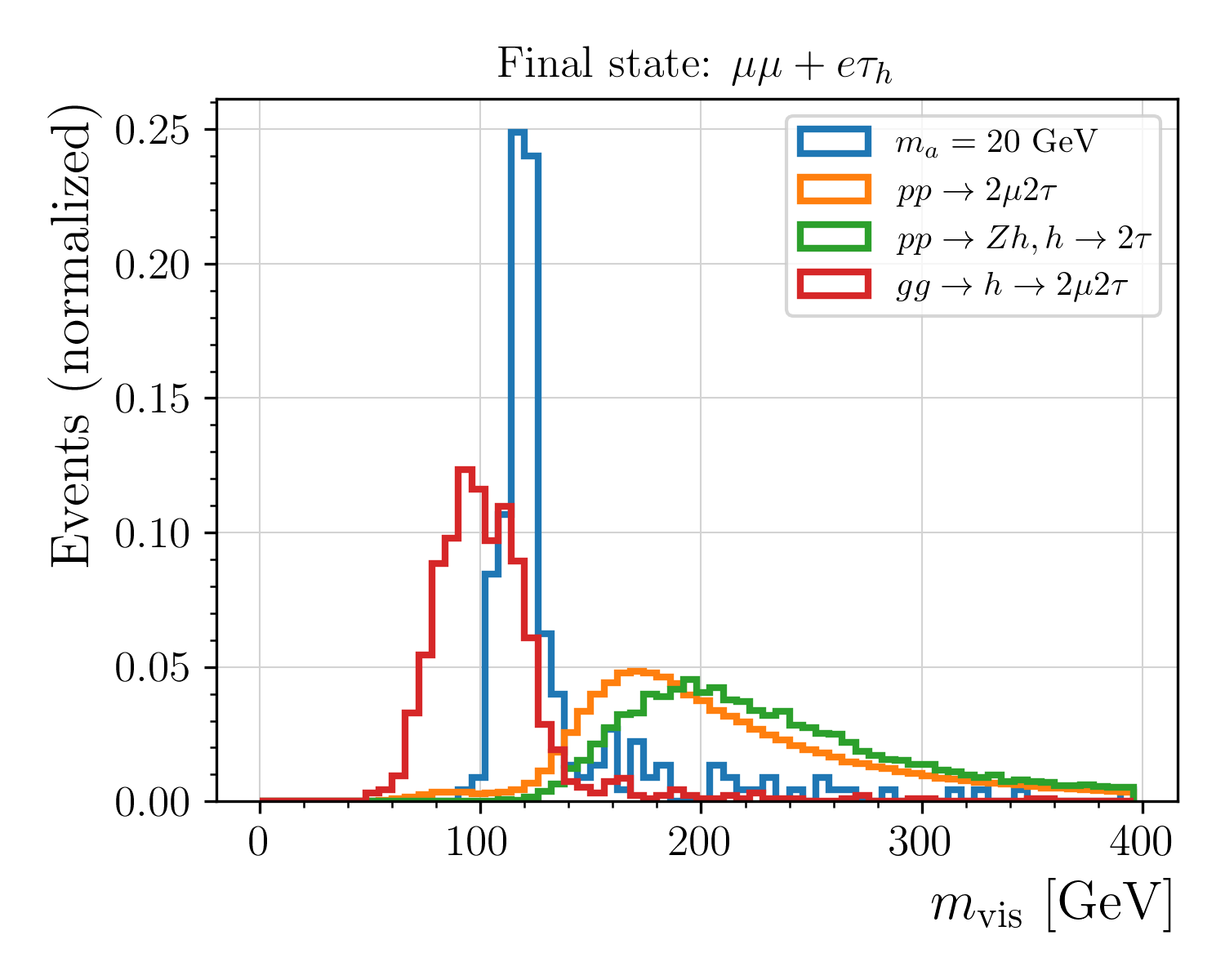}
    \includegraphics[width=0.495\linewidth]{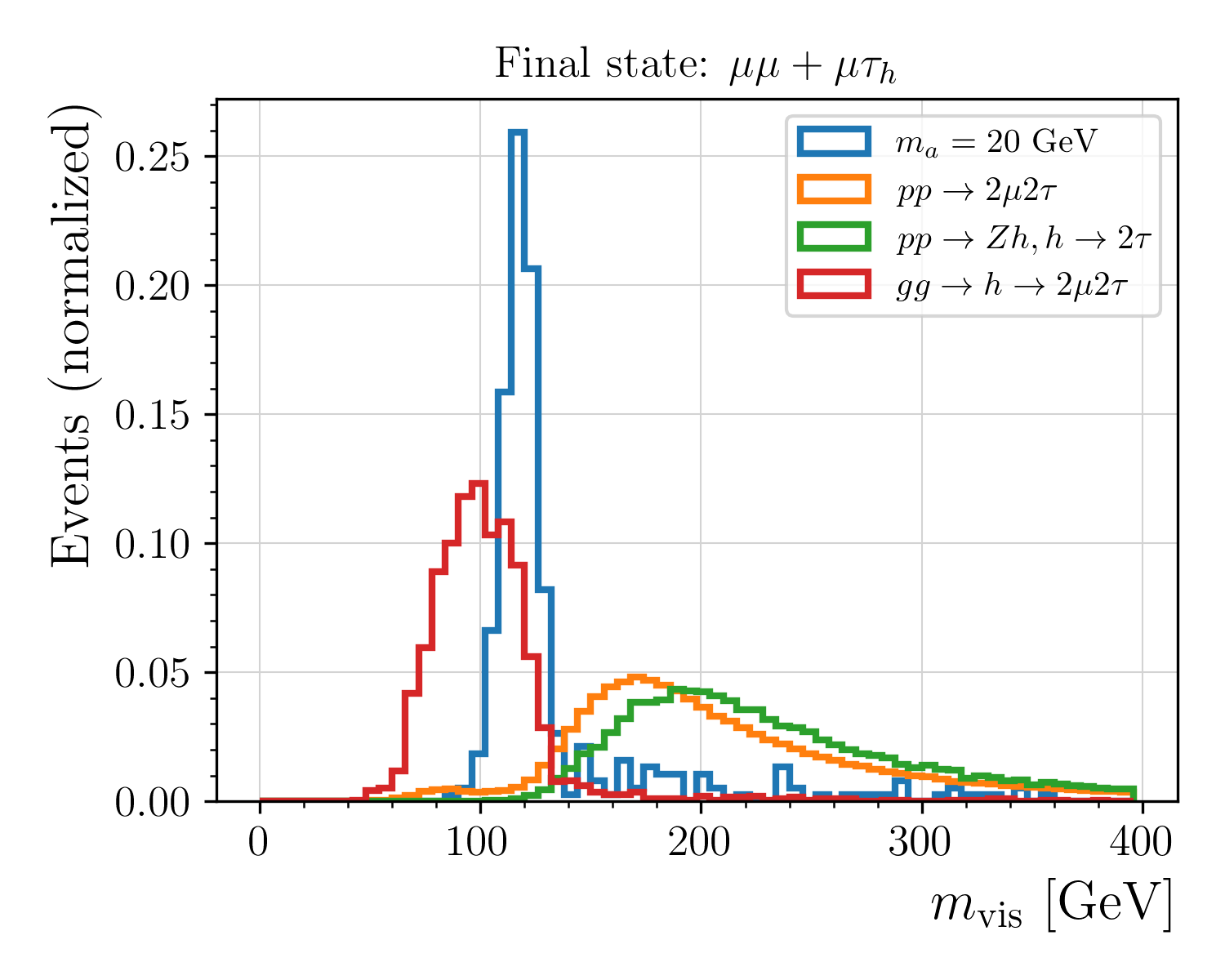}
    \includegraphics[width=0.495\linewidth]{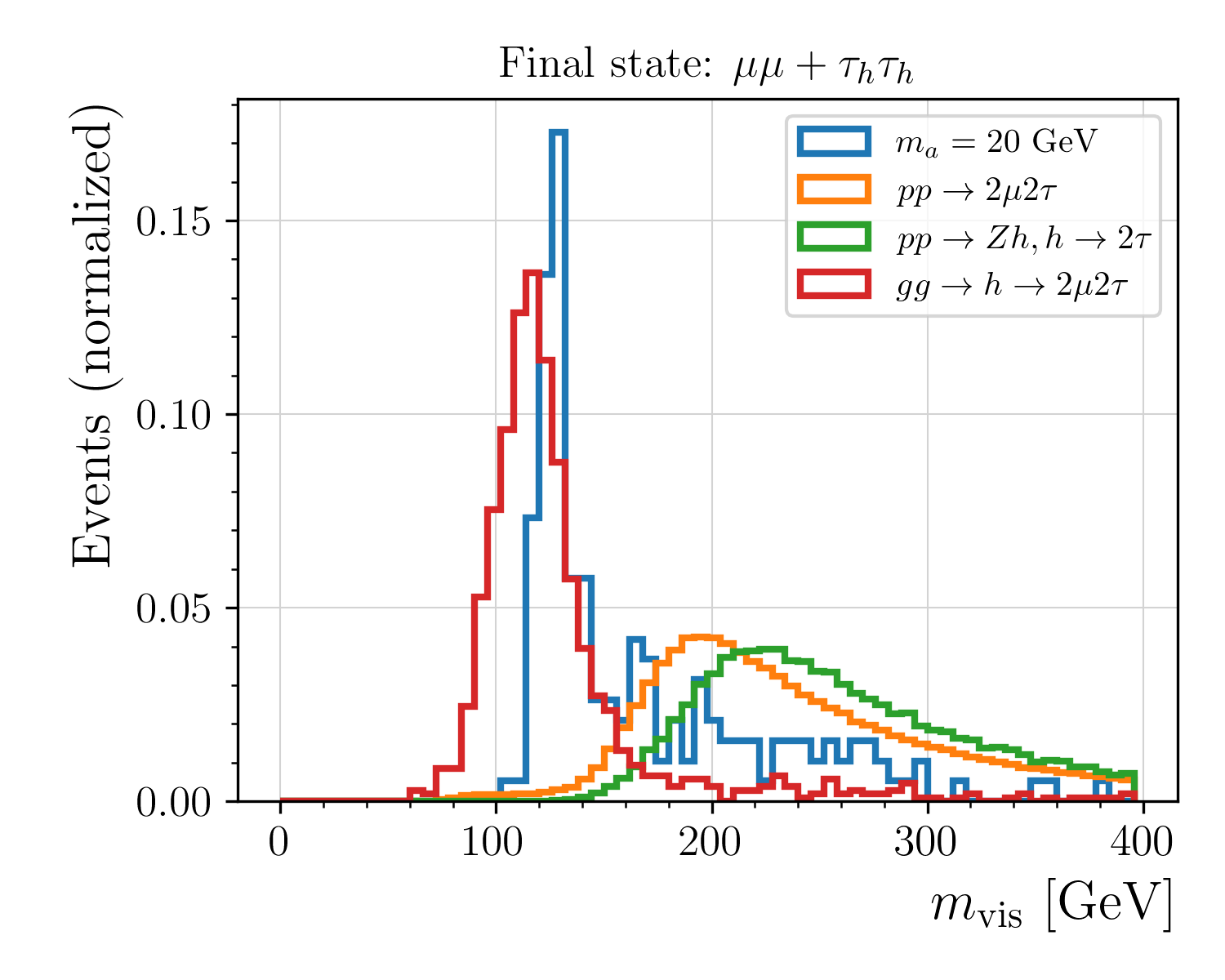}
    
    \caption{Normalized $m_{\rm vis}$ distribution for the signal and the three dominant SM backgrounds after the baseline final-state selection criteria and before the cut on $m_{\rm vis}$ (see Table~\ref{tab:cutflow1}), for the four $\tau\tau$ final states $e \mu$ (top-left), $e \tau_{h}$ (top-right) $\mu \tau_h$ (bottom-left) and $\tau_{h} \tau_{h}$ (bottom-right) considered in the analysis.}
    \label{fig:mvis_Sec4}
\end{figure}

The effect of our event selection on the signal -- for a benchmark value $m_a = 20$ GeV and assuming BR$(h \to Z a) \times$ BR$(a \to \tau\tau) = 4 \times 10^{-4}$ -- and dominant SM backgrounds is shown in Table~\ref{tab:cutflow1}.
For the signal, we have used the N$^{3}$LO value for the gluon-fusion production cross-section of the Higgs boson~\cite{Anastasiou:2016cez} at $\sqrt{s}=14$ TeV LHC, given by $\sigma( p p \to h) = 54.67$ pb. We re-stress that the number of expected SM background events after all cuts in Table~\ref{tab:cutflow1} for the subdominant backgrounds $p p \to \mu \mu \tau \tau + X$ (with $X=2 \ell,\, 2 \nu$) and $p p \to t\bar{t}Z$ (with $Z \to \mu \mu$) is $< 0.1$ at the HL-LHC with 3000 fb$^{-1}$ of integrated luminosity, and we safely disregard them in our analysis. 
In Figure~\ref{fig:mvis_Sec4} we show the normalized $m_{\rm vis}$ distribution for the signal and the dominant SM backgrounds after the baseline final-state selection criteria and before the cut on $m_{\rm vis}$ (see Table~\ref{tab:cutflow1}), for the four $\tau\tau$ final states $e \mu$, $e \tau_{h}$, $\mu \tau_h$ and $\tau_{h} \tau_{h}$ considered in the analysis. As can be seen from 
both Figure~\ref{fig:mvis_Sec4} and Table~\ref{tab:cutflow1}, the cut on $m_{\rm vis}$ renders the $p p \to Z h$ SM background negligible, and leaves $p p \to Z Z^{(*)}$ and $g g \to h$, $h \to \mu\mu \tau\tau$ as the only relevant -- and comparable -- SM backgrounds (note that the normalized nature of the distributions in Figure~\ref{fig:mvis_Sec4} may convey a different, misleading impression). In addition, as shown in Table~\ref{tab:cutflow1} the signal has a high efficiency in passing all cuts, with the exception of the baseline final-state definition after online and offline muon triggers. The strong signal reduction at this stage of the event selection is due mainly to the very low efficiency of $\tau\tau$ final state reconstruction for the $h\to Z a$, $a\to \tau\tau$ process in our analysis. Potential improvements in this efficiency at the experimental level could further increase the sensitivity of our proposed search.

\begin{figure}
    \centering
    \includegraphics[width=0.45\linewidth]{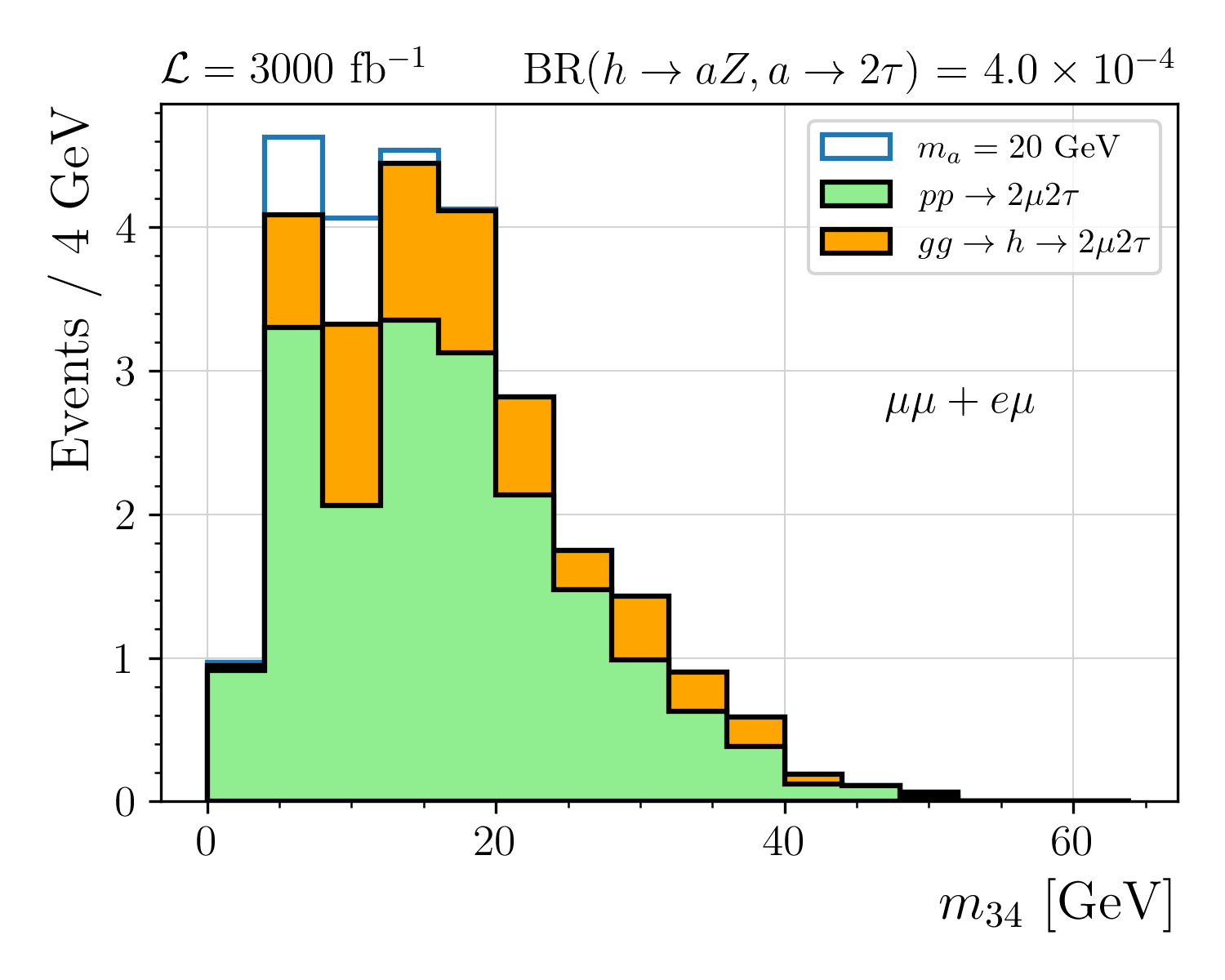}
    \includegraphics[width=0.45\linewidth]{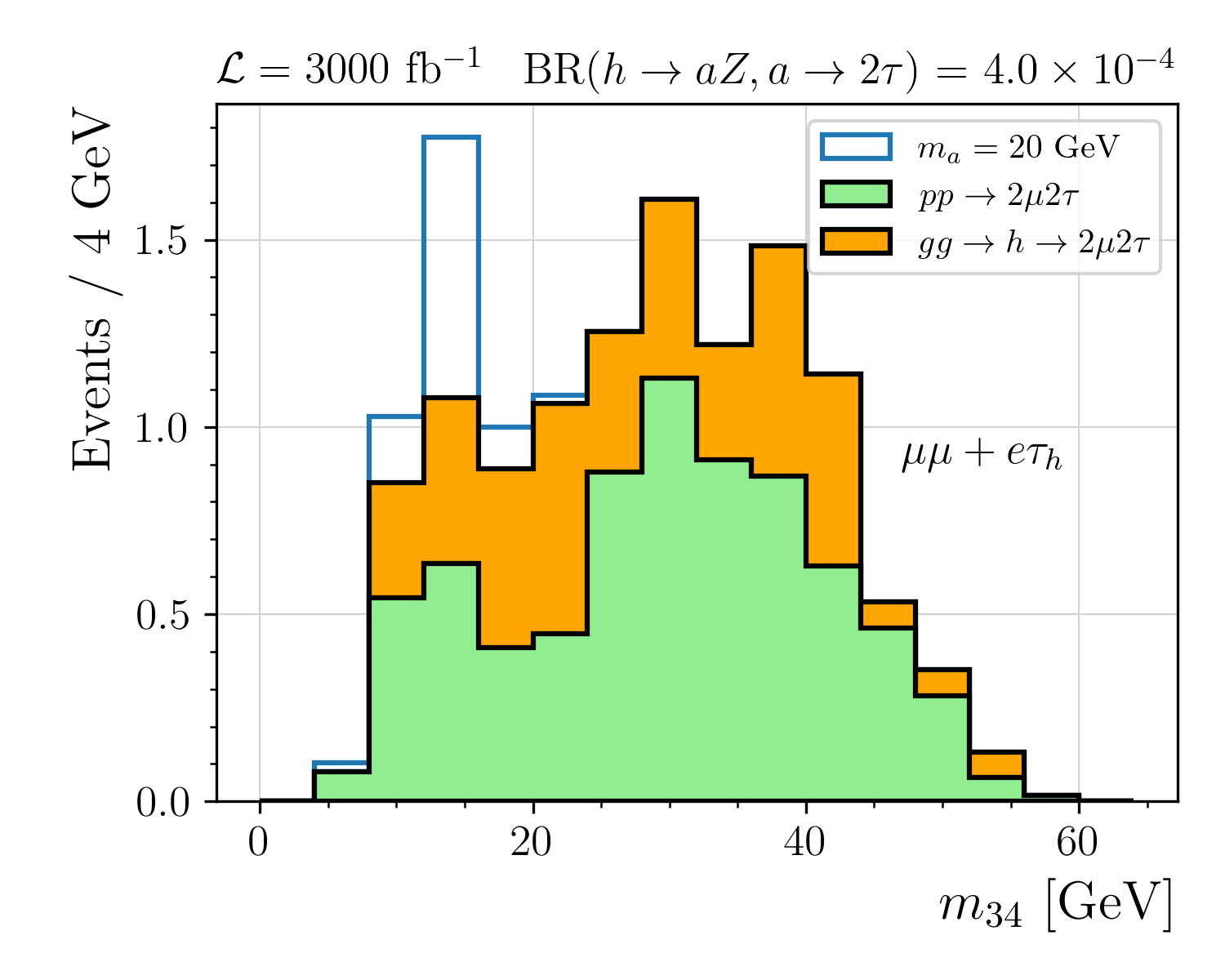}
    \includegraphics[width=0.45\linewidth]{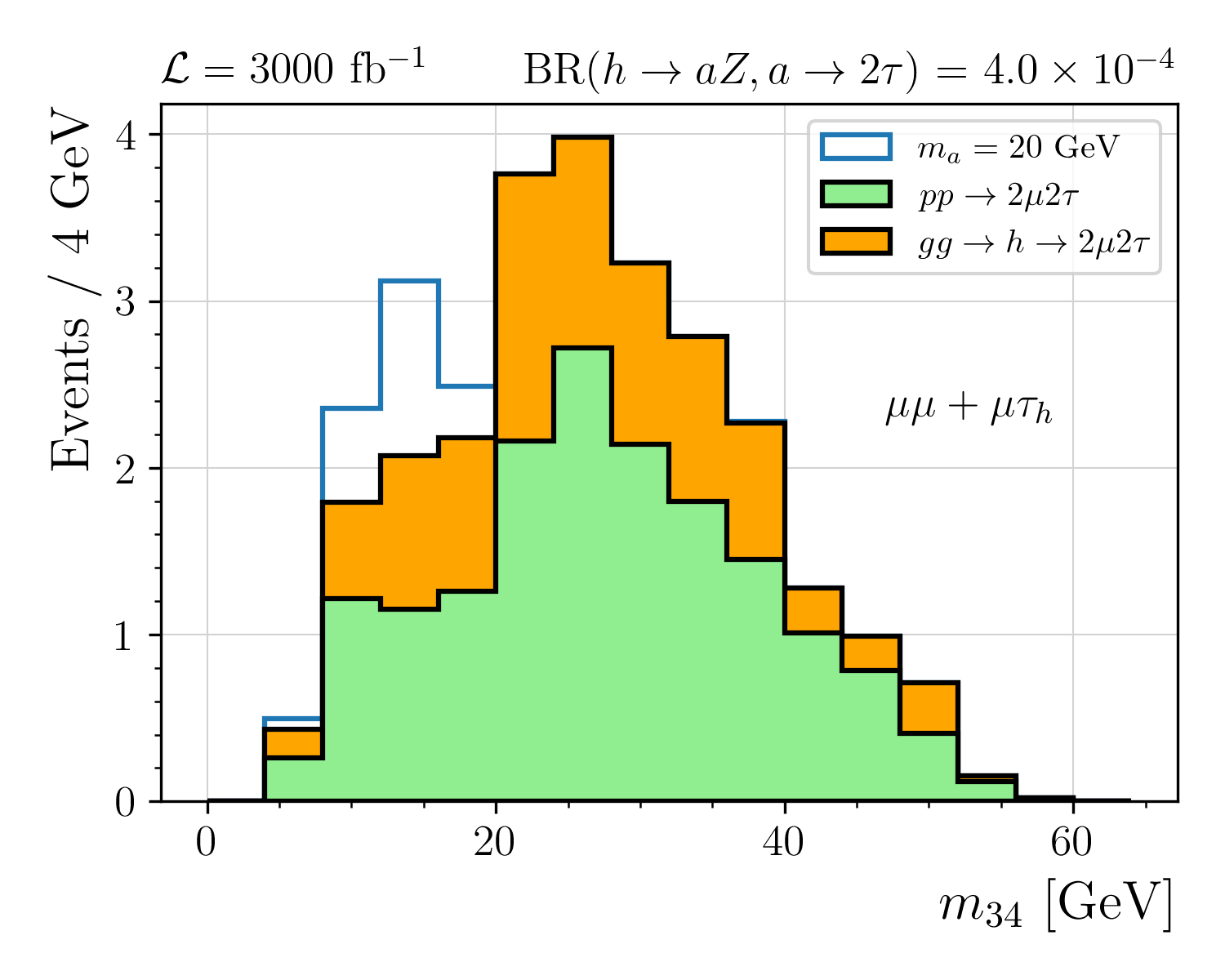}
    \includegraphics[width=0.45\linewidth]{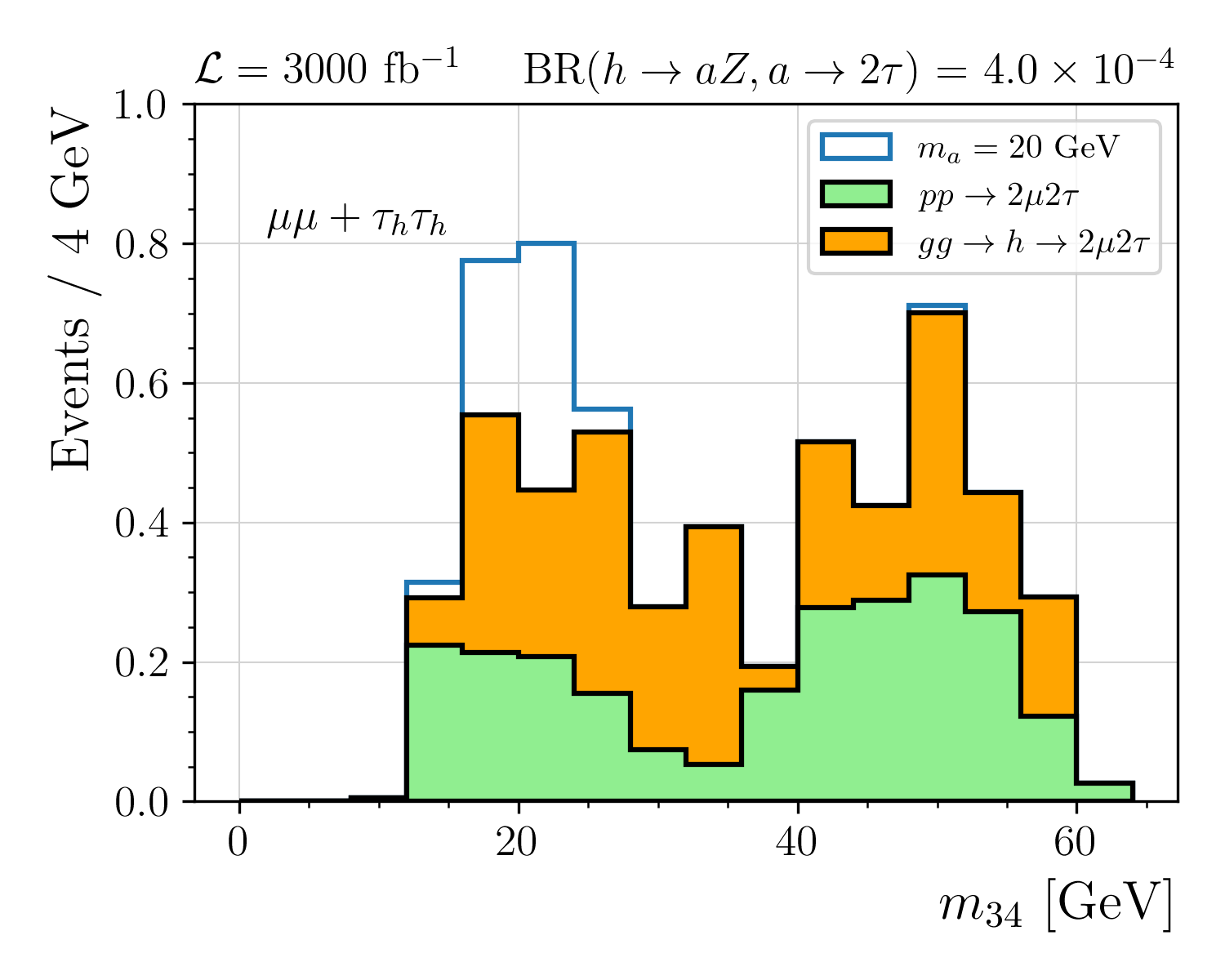}
    
    \caption{$p p \to h \to Z a \to \mu\mu\, \tau\tau$ signal (with $m_{a} = 20$ GeV) and dominant SM backgrounds (SM Higgs production through gluon-fusion $p p \to h \to  Z Z^{*}$ (with $Z Z^{(*)} \to \mu \mu \tau \tau$) and $p p \to Z Z^{(*)} \to \mu \mu \tau \tau$). Other backgrounds  (e.g. $p p \to Z h$) give a negligible contribution and are not shown.  Signal for BR$(h \to Z a) \times$ BR$(a \to \tau\tau) = 4 \times 10^{-4}$. }
    \label{fig:m34_sec4}
\end{figure}

\begin{figure}
    \centering
    \includegraphics[width=0.7\linewidth]{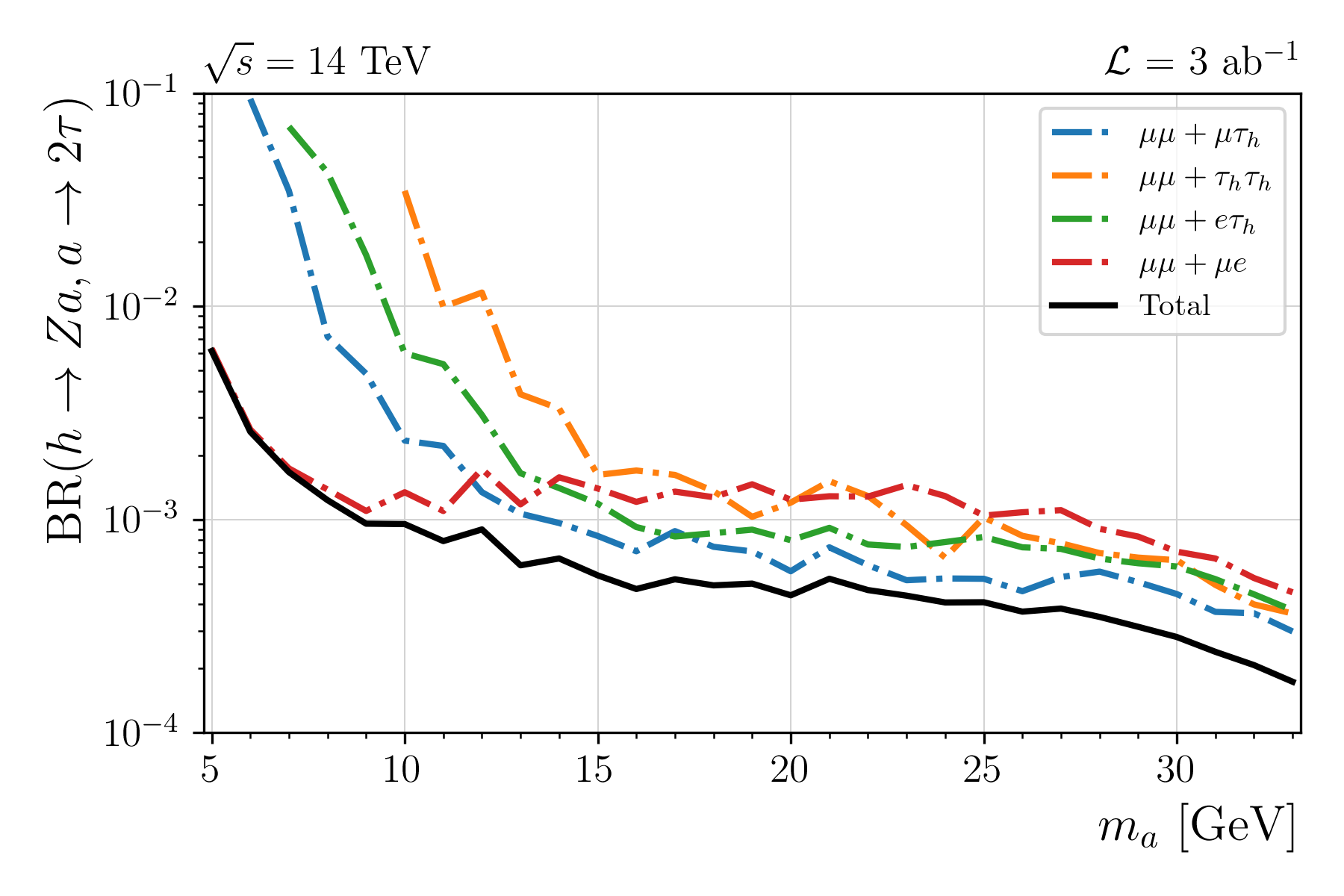}

\vspace{-4mm}
    
    \includegraphics[width=0.7\linewidth]{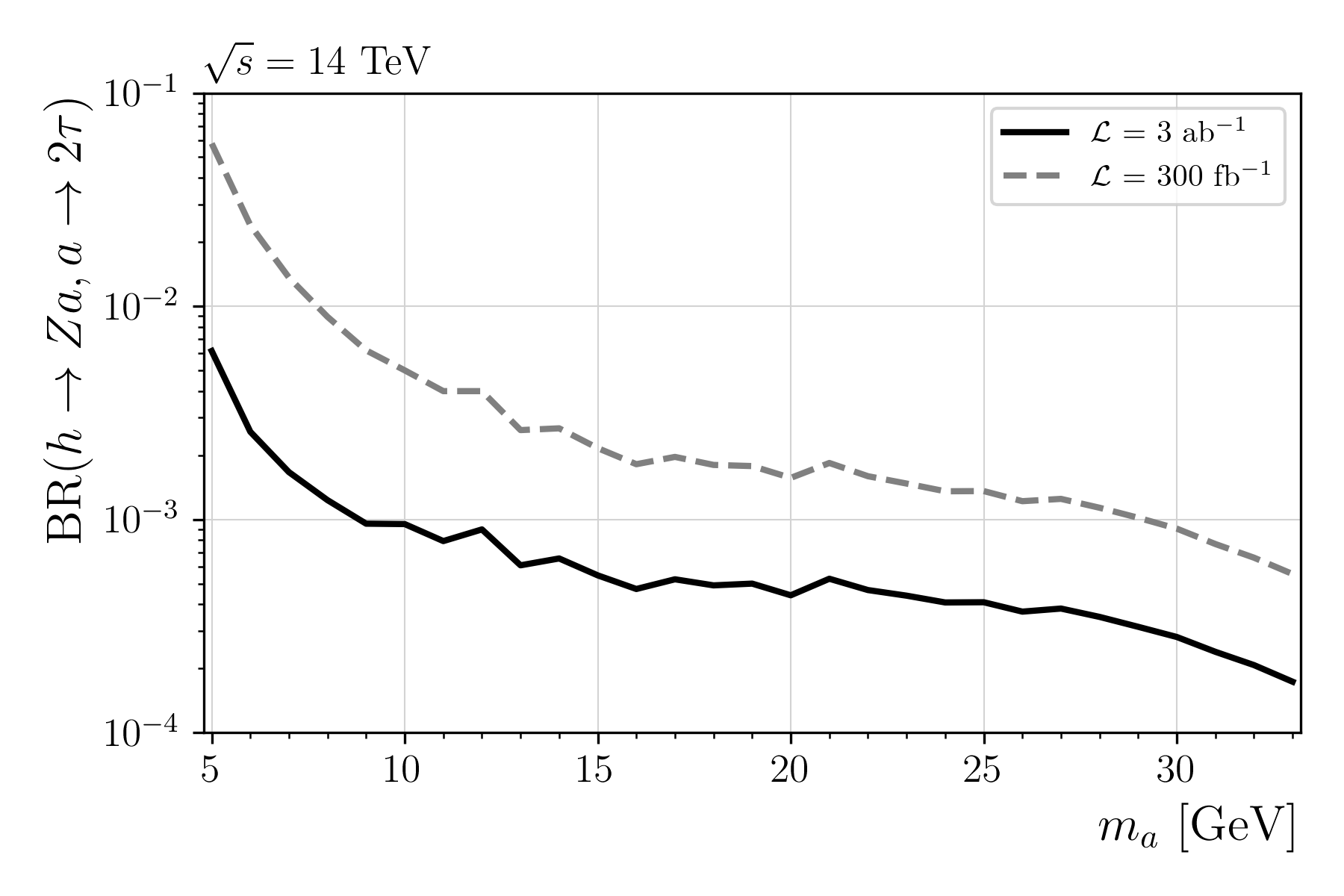}

\vspace{-7mm}
        
    \caption{Top: Projected 95$\%$ C.L. sensitivities on the BR$(h\to Za, a\to2\tau)$ for the HL-LHC (with $\sqrt{s} =$ 14 TeV and $\mathcal{L} =$ 3 ab$^{-1}$) as a function of $m_a$, for $\mu\mu + e \mu$ (red), $\mu\mu + e \tau_{h}$ (green), $\mu\mu + \mu \tau_h$ (blue), $\mu\mu + \tau_{h} \tau_{h}$ (orange) final states, and their total combination (solid black). Bottom: Comparison between the total projected 95$\%$ C.L. sensitivities on the BR$(h\to Za, a\to2\tau)$ for $\mathcal{L} =$ 300 fb$^{-1}$ (dashed grey) and $\mathcal{L} =$ 3 ab$^{-1}$ (solid black).}
    \label{fig:limits_sec4}
\end{figure}

The number of signal and SM background events in the $m_{\tau\tau}^{\rm vis} \equiv m_{34}$ kinematic variable after all analysis cuts for each of the final states $e \mu$, $e \tau_{h}$, $\mu \tau_h$ or $\tau_{h} \tau_{h}$ is depicted in Figure~\ref{fig:m34_sec4}. We then do a binned-likelihood analysis -- analogous to the one performed in Section~\ref{sec:AppI}, using the likelihood function defined in~\eqref{likelihood_NS} -- in $m_{34}$, following the binning depicted in Figure~\ref{fig:m34_sec4}. This way, we obtain the 95\% C.L. sensitivity of our analysis on BR$(h \to Z a) \times$ BR$(a \to \tau\tau)$ for the four final states $e \mu$, $e \tau_{h}$, $\mu \tau_h$ and $\tau_{h} \tau_{h}$. These 95\% C.L. sensitivities are shown in Figure~\ref{fig:limits_sec4} (top), together with their combination, for HL-LHC with 3 ab$^{-1}$ of integrated luminosity. In Figure~\ref{fig:limits_sec4} (bottom) we depict the comparison between the combined limit for HL-LHC and that obtained assuming 300 fb$^{-1}$ of integrated luminosity.\footnote{In this case, we have found that a tighter signal selection (e.g. through a narrower $m_{\mu\mu}$ window) may yield too few SM background events in the $m_{34}$ signal region for our binned-likelihood analysis to yield a robust result, which justifies the choice 50 GeV $ < m_{\mu\mu} < 106$ GeV as discussed above.} In both cases, we observe a rather strong degrading of the sensitivity for $m_a < 10$ GeV, due to the very low signal efficiencies in that mass range.

\vspace{2mm}

\begin{figure}
    \centering
    \includegraphics[width=.485\linewidth]{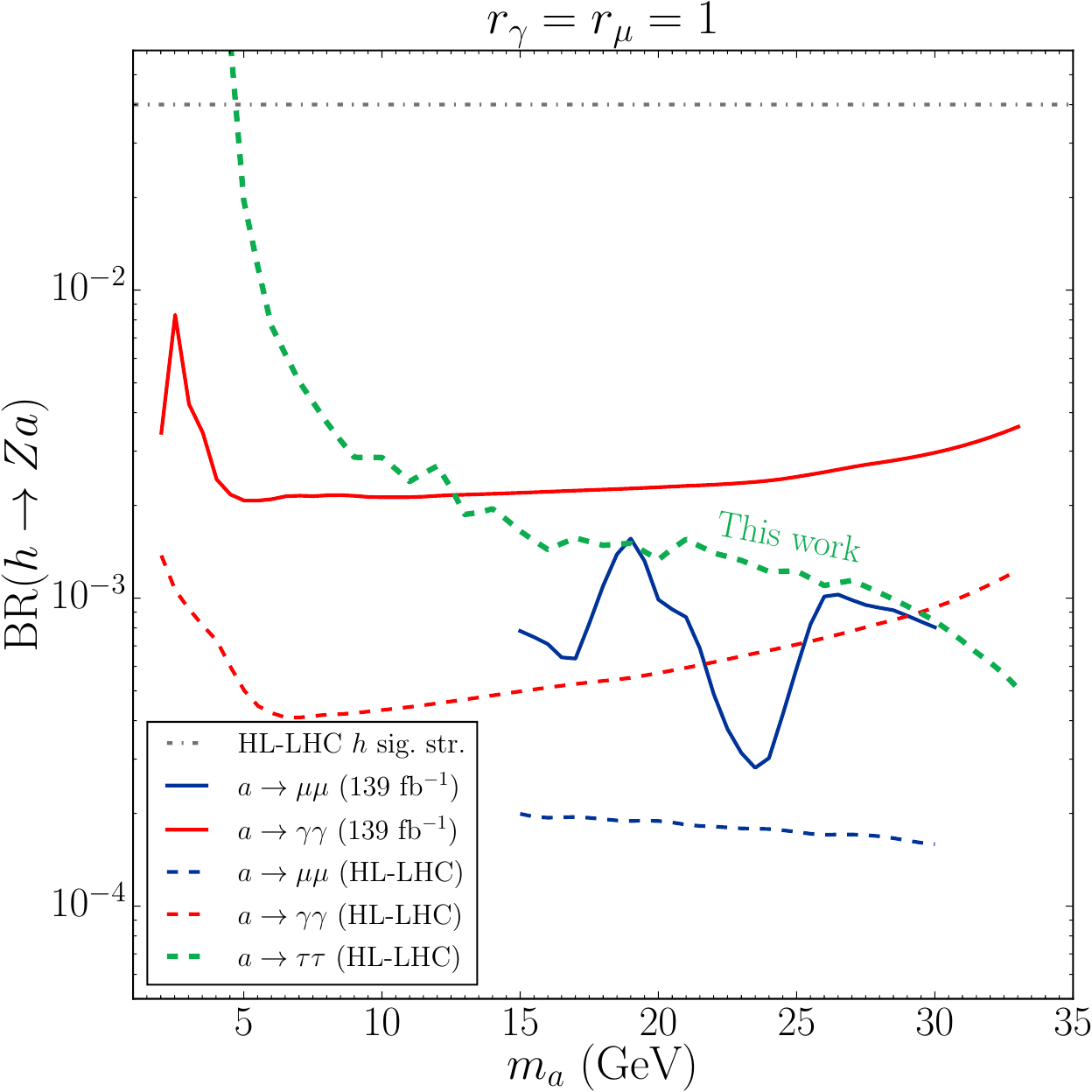}
    \hspace{1mm}
    \includegraphics[width=.485\linewidth]{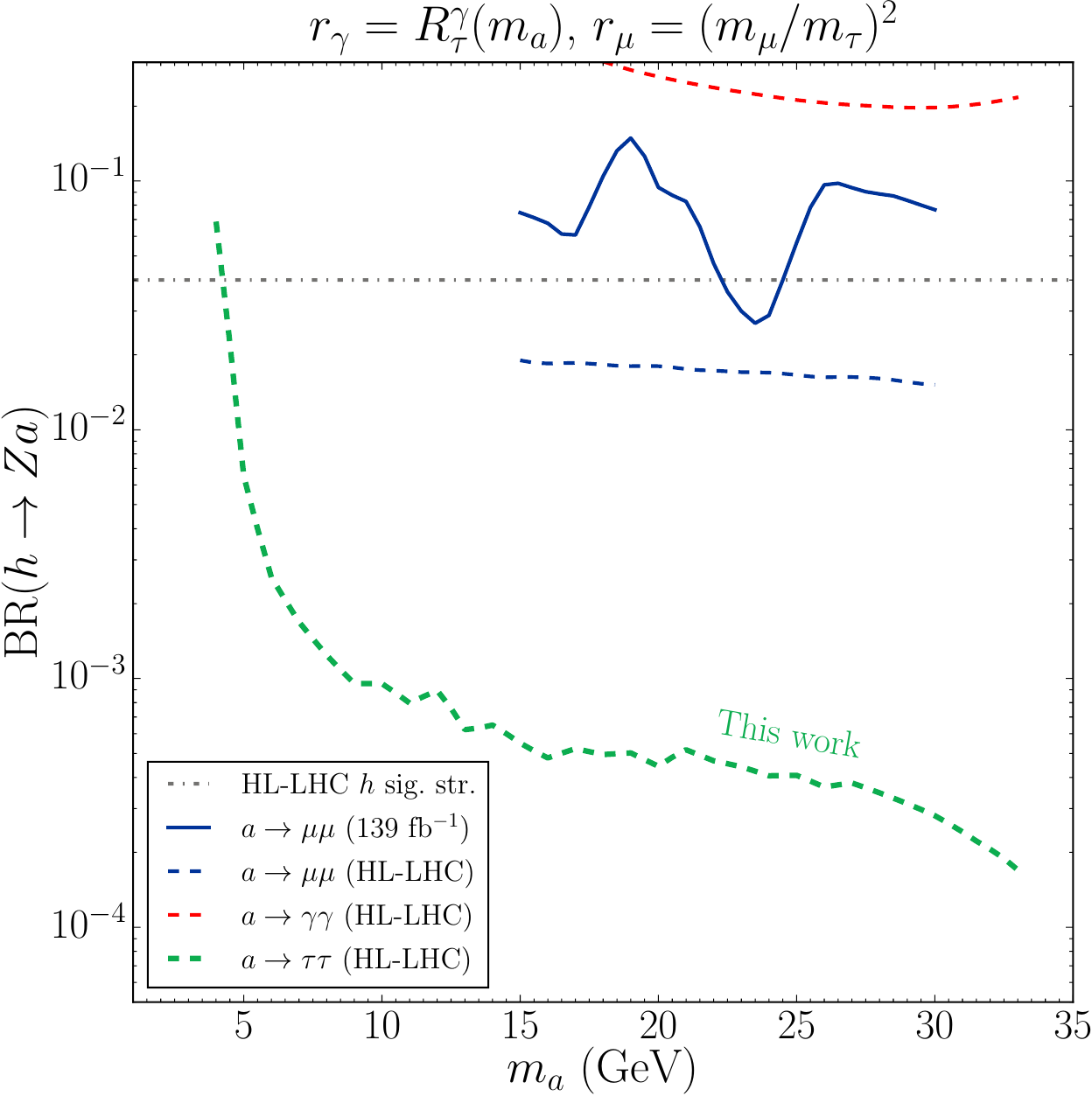}
        \caption{95\% C.L. expected limits on BR$(h \to Z a)$ as a function of $m_a$ from present ATLAS $a\to \gamma\gamma$ (solid red) and $a\to \mu\mu$ (solid blue) searches, HL-LHC (with 3 ab$^{-1}$) $a\to \gamma\gamma$ (dashed red) and $a\to \mu\mu$ (dashed blue) projections, and our $a\to \tau\tau$ sensitivity projection (dashed green), for $r_{\mu} = r_{\gamma} = 1$ (left) and $r_{\mu} = m_{\mu}^2/m_{\tau}^2$, $r_{\gamma} = R_{\tau}^{\gamma} (m_a)$ (right). The dash-dotted grey line corresponds to the expected HL-LHC sensitivity from a global fit to $h$ data (see text for details).
        }
    \label{fig:otherdecays}
\end{figure}

We can now compare the expected sensitivity of our proposed $h\to Z a$, $a\to \tau \tau$ search with that of existing searches for $h\to Z a$ exotic Higgs decays in $a \to \mu \mu$ final states (the ATLAS search~\cite{ATLAS:2021ldb} discussed in Section~\ref{sec:AppI}) and $a \to \gamma \gamma$~\cite{ATLAS:2023etl}. Since both experimental searches are performed for $\mathcal{L} =$ 139 fb$^{-1}$ of integrated luminosity, we naively extrapolate their current \emph{expected} 95\% C.L. limits to the HL-LHC with $\mathcal{L} =$ 3000 fb$^{-1}$, to perform a direct comparison of sensitivities.\footnote{We note that our analysis is performed at $\sqrt{s} = 14$ TeV, as opposed to the experimental $a \to \mu \mu$ and $a \to \gamma \gamma$ searches, performed for $\sqrt{s} = 13$ TeV. The effect in the comparison should however be minor.} We thus rescale the current experimental limits by $\sqrt{\mathcal{L}_{\rm current}/\mathcal{L}_{\rm HL-LHC}} = \sqrt{139/3000} \sim 0.215$. Considering for simplicity that only the three decay modes $a\to \tau \tau$, $a\to \mu \mu$ and $a \to \gamma \gamma$ are present (otherwise we would just rescale the sensitivities for the three channels by a common factor), i.e. ${\rm BR} (a \to \tau \tau) + {\rm BR} (a \to \mu \mu) + {\rm BR} (a \to \gamma \gamma) = 1$,  we show in Figure~\ref{fig:otherdecays} the current bounds and expected HL-LHC sensitivities to ${\rm BR} (h\to Z a)$ for different assumptions on the ratios $r_{\gamma}$ and $r_{\mu}$ 
\begin{equation}
r_{\gamma} \equiv \frac{{\rm BR} (a \to \gamma \gamma) } {{\rm BR} (a \to \tau \tau)} \, \quad , \quad r_{\mu} \equiv \frac{{\rm BR} (a \to \mu \mu)} {{\rm BR} (a \to \tau \tau)} \, .
\end{equation}
We have also included in Figure~\ref{fig:otherdecays} the 4\% expected HL-LHC sensitivity from a global fit to $h$ data~\cite{deBlas:2019rxi}, under the assumption that current theoretical systematic uncertainties are halved and experimental uncertainties are reduced to reach projected HL-LHC values. In Figure~\ref{fig:otherdecays} (left) we assume $r_{\gamma} = r_{\mu} = 1$, case in which the existing searches in $a \to \mu \mu$ and $a \to \gamma \gamma$ final states are more sensitive than our proposed search, given the significantly higher efficiency of muon and photon reconstruction over $\tau$-leptons at the LHC. Yet, in general such values for $r_{\gamma}$ and $r_{\mu}$ are not realized in BSM models, but rather one expects $r_{\mu} \ll 1$ and/or $r_{\gamma} \ll 1$. Specifically, in many BSM scenarios one finds $r_{\mu} \simeq m_{\mu}^2/m_{\tau}^2 \sim 3.6 \times 10^{-3}$ -- recall Eq.~\eqref{eq:BR_Hierarchy} --, just as the ratio $\Gamma (h \to \mu\mu)/\Gamma (h \to \tau\tau)$ in the SM. Similarly, in the SM the ratio $\Gamma (h \to \gamma\gamma)/\Gamma (h \to \tau\tau) \equiv R_{\tau}^{\gamma} (m_h) \ll 1$ for every value of the Higgs boson mass $m_h$. In Figure~\ref{fig:otherdecays} (right) we then show the corresponding sensitivities to ${\rm BR} (h\to Z a)$ for $r_{\mu} = m_{\mu}^2/m_{\tau}^2$ and $r_{\gamma} = R_{\tau}^{\gamma} (m_a)$, mimicking the SM ratios (a ``\emph{SM-like ALP}" $a$). It is apparent that for such a SM-like $a$ particle the $h\to Z a$, $a\to \tau\tau$ search becomes by far the most sensitive probe of this exotic Higgs decay.

In the next section, we discuss this interplay among different searches in the context of a specific BSM model, an ALP extension of the SM. We demonstrate that our proposed search can be used to probe otherwise inaccessible and well-motivated -- in the context of the muon $g-2$ anomaly -- regions of the parameter space of the model.
\section{\texorpdfstring{\textbf{Accessing the $\bm{(g-2)_{\mu}}$ parameter space of ALPs via exotic Higgs decays}}{Accessing the (g-2)_mu parameter space of ALP models}}
\label{sec:models}

At present, the muon $g-2$ anomaly is given by 
\begin{equation}
   \delta a_{\mu} \equiv a_{\mu}^{\rm exp} - a_{\mu}^{\rm SM} = (24.9 \pm 4.8) \times 10^{-10} \, ,
   \label{muong-2data}
\end{equation}
with the most recent experimental value~\cite{Muong-2:2023cdq} having been determined in 2023 by the \texttt{Muon g-2} Collaboration at Fermilab, and the SM prediction computed in~\cite{Aoyama:2020ynm}. While recent lattice field theory results~\cite{Borsanyi:2020mff,Ce:2022kxy} have put in question the BSM origin of such anomaly -- rather putting the focus on the determination of~$a_{\mu}^{\rm SM}$, which involves the use of data-driven techniques~\cite{ExtendedTwistedMassCollaborationETMC:2022sta} --, if one takes Eq.~\eqref{muong-2data} seriously it represents almost a $5.2\, \sigma$ discrepancy with the SM.

In this section we consider the muon $g-2$ anomaly at face value, and explore a possible BSM solution to the muon $g-2$ anomaly via a light ALP $a$ that couples to SM leptons, photons and the 125 GeV Higgs boson. In such a setup -- which has been discussed before in~\cite{Bauer:2017ris} -- , we show below that our proposed exotic Higgs decay $h \to Z a \to \mu\mu\, \tau\tau$ search can probe part of the ALP parameter space region that explains the $(g-2)_{\mu}$ anomaly, in strong complementarity with existing LHC experimental searches of the $h\to Z a$ exotic Higgs decay in the $a \to \gamma\gamma$ final state~\cite{ATLAS:2023etl}.

\begin{figure}[h!]
    \centering
    \includegraphics[width=.35\linewidth]{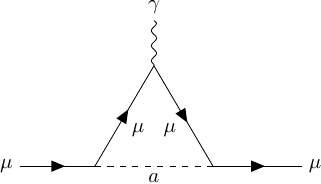}
\hspace{1cm}
        \includegraphics[width=.35\linewidth]{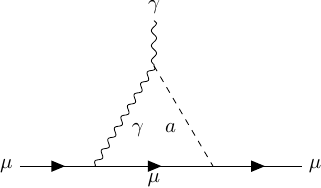}
    \caption{1-loop Feynman diagrams contributing to the $g-2$ of the muon via an ALP coupled to photons and SM leptons.}
    \label{fig:Feynman_g-2}
\end{figure}

 We start our discussion by introducing the effective field theory (EFT) operators which couple an ALP to photons and SM leptons, respectively (a detailed discussion of the effective Lagrangian of an ALP coupled to the SM can e.g.~be found in~\cite{Bauer:2017ris,Brivio:2017ije}): 
\begin{equation}
\label{eq:ALP_1}
\mathcal{L}_{\rm ALP} \supset \frac{c_{\gamma\gamma}}{f_a} a\, F_{\mu\nu}  \tilde{F}^{\mu\nu} + \frac{c_{\ell\ell}}{f_a} \, \sum_{\ell}  (\partial^\mu a) \, \bar{\ell} \gamma_{\mu} \gamma_{5} \ell
\end{equation}
with $c_{\gamma\gamma}$ and $c_{\ell\ell}$ the respective Wilson coefficients and $f_a$ the EFT cut-off scale. We have assumed in~\eqref{eq:ALP_1} that the SM leptons $\ell = e,\, \mu,\, \tau$ couple to the ALP $a$ via a single (universal) Wilson coefficient $c_{\ell\ell}$ (rather than via different coefficients for the different SM lepton flavours). Such an ALP $a$ induces a BSM contribution to the $g-2$ of the muon which is given at 1-loop via the two diagrammatic contributions shown in Figure~\ref{fig:Feynman_g-2}. To compute those, we follow~\cite{Bauer:2017ris} and obtain\footnote{Following the phenomenological analysis of~\cite{Bauer:2017ris}, here we have neglected a possible short-distance contribution, as well as a contribution from a coupling of the ALP to a photon and a $Z$ boson (which we do not include in~\eqref{eq:ALP_1}) via a $c_{Z\gamma}$ Wilson coefficient.}
\begin{equation}
\delta a_{\mu} \simeq  - m_{\mu}^2 \left\lbrace \left(\frac{c_{\ell\ell}}{4 \pi f_a} \right)^2 F_1 \left(\frac{m_a^2}{m_{\mu}^2}\right) + \frac{2\,\alpha_{\rm EM}}{\pi} \frac{c_{\ell\ell}\, c_{\gamma\gamma}}{f_a^2} \left[ {\rm ln} \left(\frac{Q^2}{m_{\mu}^2}\right) + \delta_2 + 3 - F_2 \left(\frac{m_a^2}{m_{\mu}^2}\right) \right] \right\rbrace \, .
\end{equation}
The loop functions $F_1(x)$ and $F_2(x)$ are given by:
\begin{eqnarray}
F_1(x) & =& 1 +2\,x + x(1-x) {\rm ln} (x)  - 2\, x (3-x) \sqrt{\frac{x}{4-x}}\, {\rm arccos} (\sqrt{x}/2)  \\
F_2(x) & = & 1 -\frac{x}{3} + \frac{x^2}{6} {\rm ln} (x) + \frac{2+x}{3}  \sqrt{x(4-x)} \, {\rm arccos} (\sqrt{x}/2) 
\end{eqnarray}
while the scheme-dependent constant is fixed to $\delta_2 = -3$.

\begin{figure}[h]
    \centering
    \includegraphics[width=.7\linewidth]{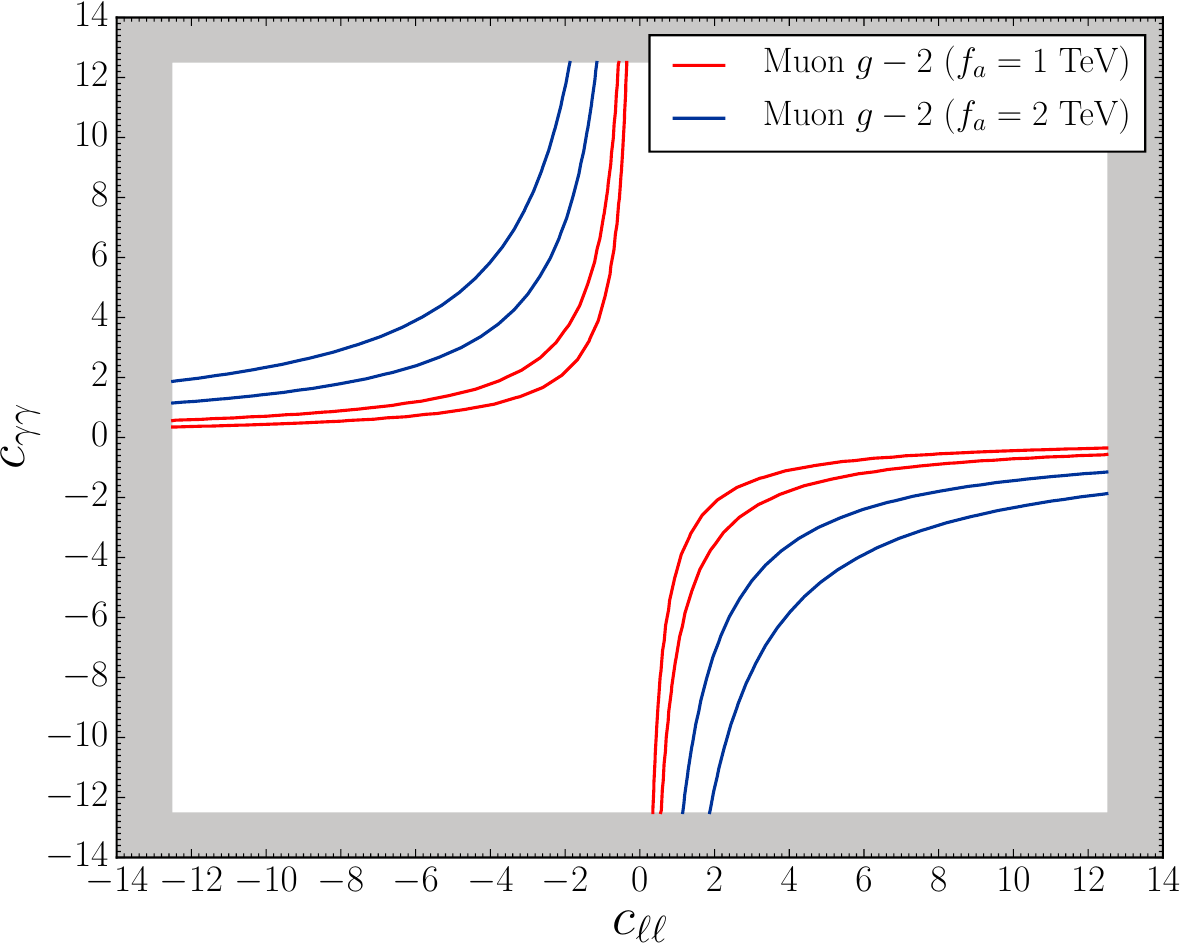}
    \caption{Region in the ($c_{\ell\ell}$, $c_{\gamma\gamma}$) plane for which the $g-2$ of the muon is fitted at the $2\sigma$ level for $f_a = 1$ TeV (red) and $f_a = 2$ TeV (blue). The grey region corresponds to either $|c_{\ell\ell}| > 4\pi $ or $|c_{\gamma\gamma}| > 4\pi $ (or both), and is beyond the naive perturbativity limit of our theory.}
    \label{fig:ALP_g-2_1}
\end{figure}

We note that the 1-loop ALP contribution to the muon $g-2$ proportional to $c_{\ell\ell}^2$ in Figure~\ref{fig:Feynman_g-2} (left) yields a BSM shift of the opposite sign of what is needed to explain the anomaly in(~\ref{muong-2data}). Such shift can be compensated by that of the 1-loop ALP contribution proportional to $c_{\ell\ell}\,c_{\gamma\gamma}$, shown in Figure~\ref{fig:Feynman_g-2} (right), if the two Wilson coefficients have opposite signs. This is explicitly seen in Figure~\ref{fig:ALP_g-2_1}, where we show the parameter space region (for a benchmark ALP mass $m_a = 20$ GeV) in the ($c_{\ell\ell}$, $c_{\gamma\gamma}$) for which the muon $g-2$ is explained at the $2\sigma$ level (for two values of the $f_a$ scale).


Turning now to the collider phenomenology of the ALP, the interactions in~\eqref{eq:ALP_1} lead to two decay modes of the ALP, into photons $a \to \gamma\gamma$ and SM leptons $a \to \ell\ell$ respectively, with the corresponding partial widths given by~\cite{Bauer:2017ris} 
\begin{equation}
\label{eq:ALP_3}
\Gamma_{\gamma\gamma} = \frac{m_a^3}{4\, \pi\, f_a^2} \,  (c_{\gamma\gamma}^{\rm eff})^2 \quad , \quad 
\Gamma_{\ell\ell} = \frac{m_a \, m_\ell^2}{8\,\pi\,f_a^2} \, (c_{\ell\ell}^{\rm eff})^2 \, \sqrt{1 - \left(\frac{2\, m_\ell}{m_a}\right)^2}
\end{equation}
where the couplings $c_{\gamma\gamma}^{\rm eff}$ and  $c_{\ell\ell}^{\rm eff}$ include 1-loop corrections~\cite{Bauer:2017ris,Bonilla:2021ufe} and are approximately given by\footnote{The complete 1-loop expressions can be found in~\cite{Bonilla:2021ufe}. } 
\begin{eqnarray}
\label{eq:ALP_4}
c_{\gamma\gamma}^{\rm eff} & \simeq &  c_{\gamma\gamma} + \frac{\alpha_{\rm EM} c_{\ell\ell}}{4\, \pi} \left[B_1\left(\frac{4 m_{\tau}^2}{m_a^2}\right)  + B_1\left(\frac{4 m_{\mu}^2}{m_a^2}\right) \right]\\
c_{\ell\ell}^{\rm eff} & \simeq & c_{\ell\ell} + \frac{\alpha_{\rm EM} c_{\gamma\gamma}}{\pi} \left[ 3\, {\rm log}\left(\frac{f_a^2}{m_\ell^2}\right) - 4 - \frac{2 \pi^2}{3} - \frac{1}{2} \left(   {\rm log}\left(\frac{m_\ell^2}{m_a^2}\right) + i\,\pi \right)^2\right]
\end{eqnarray}
with the function $B_1(x) \simeq 1$ for $x\ll 1$ (its complete expression can be found in~\cite{Bauer:2017ris}).

\vspace{2mm}

Besides the ALP interactions in~\eqref{eq:ALP_1}, we consider in this work a coupling of the ALP to the SM Higgs boson and a $Z$ boson. The specific EFT operator(s) that lead to this interaction depend on the specific assumption of a linear vs non-linear ALP Lagrangian~\cite{Brivio:2017ije}, and we do not choose here a specific realization. Instead, the partial width for the decay $h \to Z a$ can be generally parametrized in terms of an effective (dimensionless) Wilson coefficient $c_{hZa}$, which itself encodes the details of the EFT interaction between the ALP and the Higgs boson~\cite{Bauer:2017ris,Brivio:2017ije}
\begin{equation}
\label{eq:ALP_2}
\Gamma_{h\rightarrow Z a} = \frac{m_h^3}{16\pi f_a^2} c_{hZa}^2 \lambda^{3/2}
\end{equation}
with $\lambda = (1- (m_Z^2-m_a^2)/m_h^2)^2 - 4 \, m_Z^2m_a^2/m_h^4$. 

\begin{figure}[h]
    \centering
    \includegraphics[width=.49\linewidth]{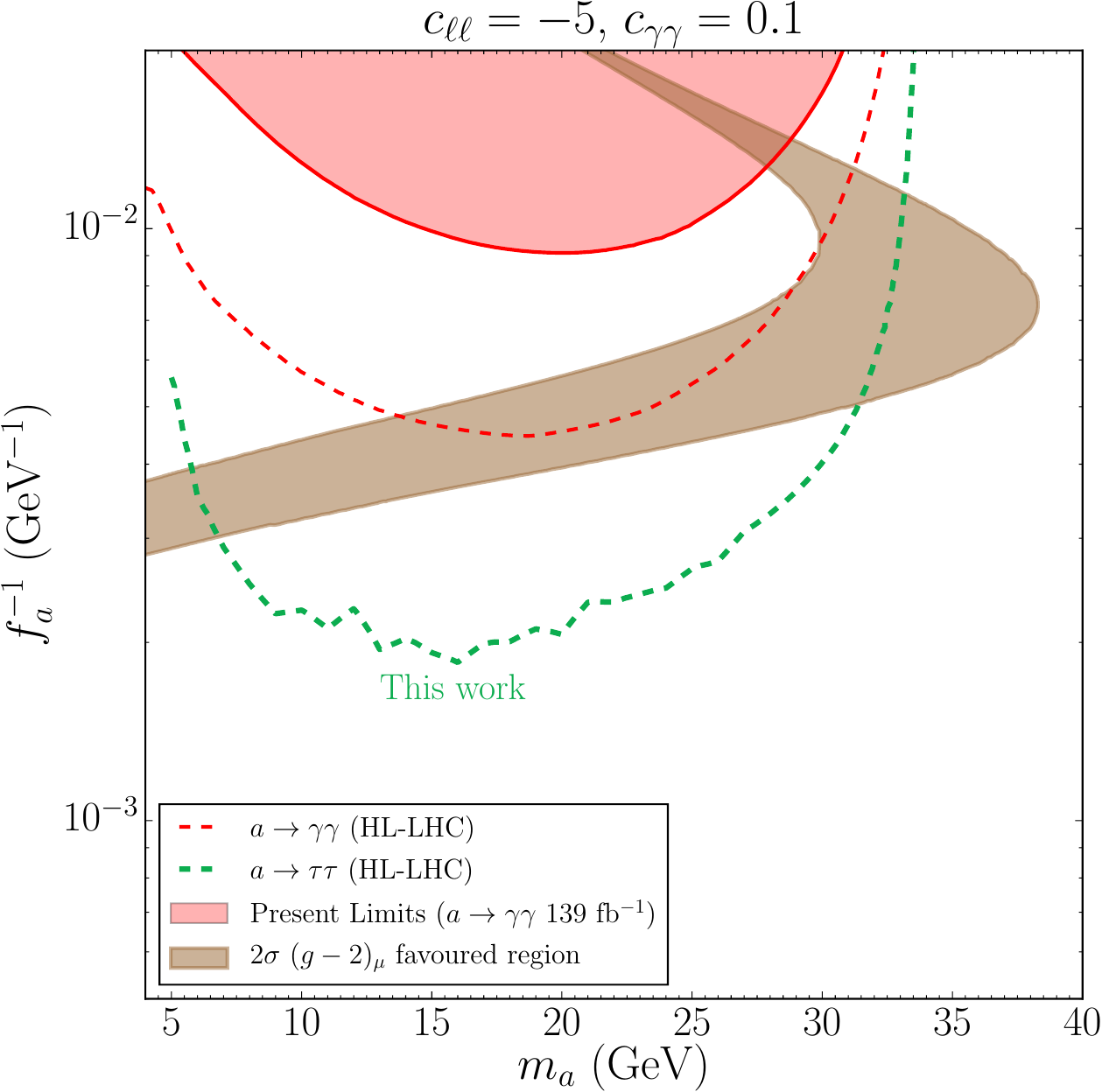}
        \includegraphics[width=.49\linewidth]{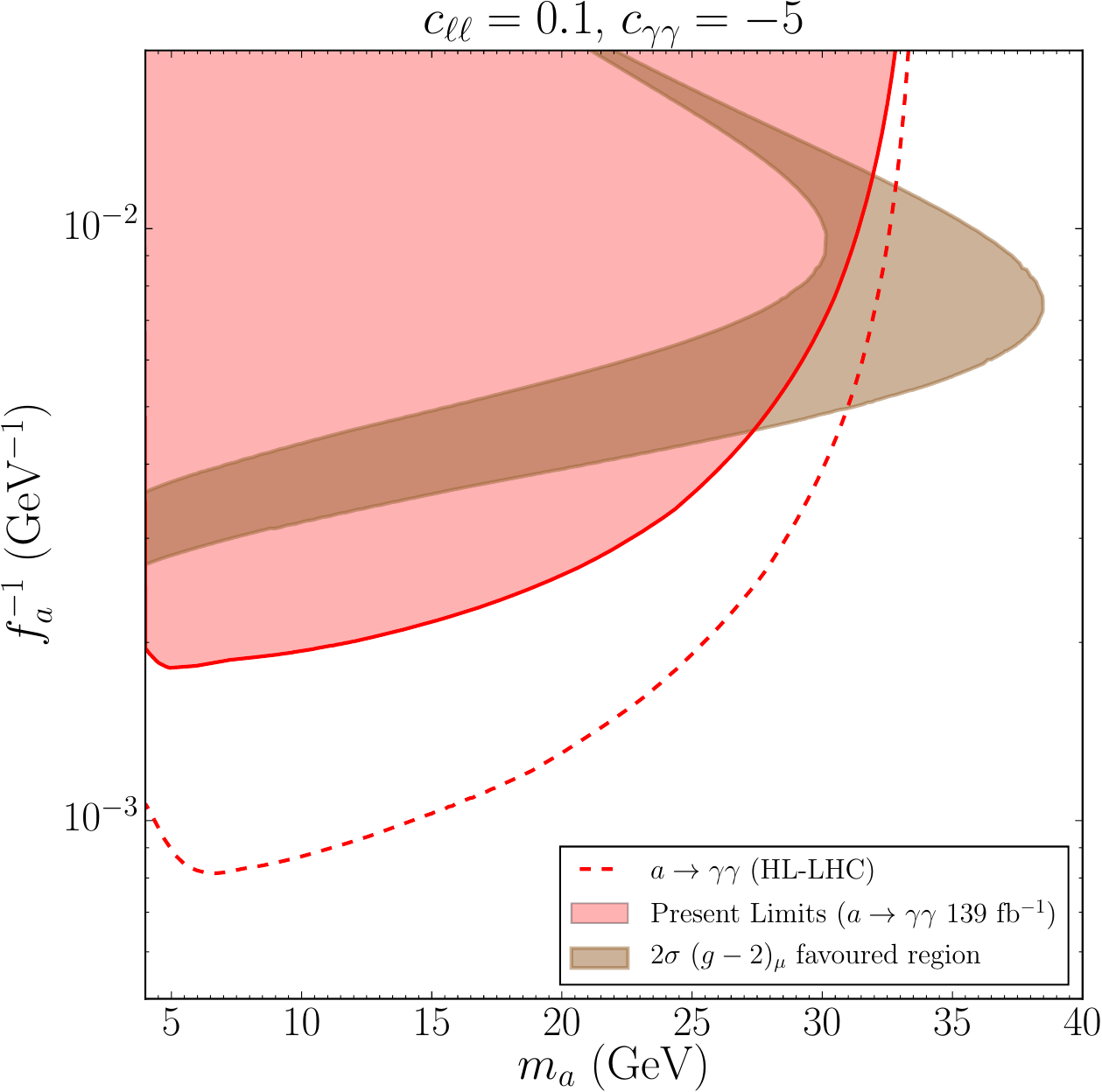}
    \caption{95\% C.L. sensitivity to $f_a^{-1}$ as a function of $m_a$, from current $h \to Z a$, $a\to \gamma \gamma$ ATLAS searches~\cite{ATLAS:2023etl} (red region), HL-LHC projections of these (dashed-red), and our proposed search $h \to Z a$, $a\to \tau\tau$ at HL-LHC (dashed green), for $c_{\ell\ell} = -5$, $c_{\gamma\gamma} = 0.1$ (left) and $c_{\ell\ell} = 0.1$, $c_{\gamma\gamma} = -5$ (right). In both cases, we fix $c_{aZh} = 0.015$. We also show the favoured  muon $g-2$ regions at the 2$\sigma$ level (brown region).}
    \label{fig:ALP_g-2_3}
\end{figure}

We can now explore the interplay between the various decay modes of the ALP $a$ towards probing the region of parameter space favoured by an explanation of the muon $g-2$ anomaly, through searches for the exotic Higgs decay $h \to Z a$. In Figure~\ref{fig:ALP_g-2_3} we fix  $c_{aZh} = 0.015$ and show the current and projected 95\% C.L. sensitivity -- in $f_a^{-1}$ -- of the existing searches for $h \to Z a$ in $a\to \gamma \gamma$ together with that of our proposed search via $a\to \tau\tau$ (the $h\to Z a$, $a\to\mu\mu$ searches are found not to yield any relevant limits, due to the very small $a\to\mu\mu$ branching fraction), for $c_{\ell\ell} = -5$, $c_{\gamma\gamma} = 0.1$ (Figure~\ref{fig:ALP_g-2_3} (left)) and $c_{\ell\ell} = 0.1$, $c_{\gamma\gamma} = -5$ (Figure~\ref{fig:ALP_g-2_3} (right)), to showcase the $|c_{\ell\ell} | \gg |c_{\gamma\gamma}|$ -- for which our $a\to\tau\tau$ search yields the strongest sensitivity -- and $|c_{\ell\ell} | \ll |c_{\gamma\gamma}|$ -- for which our search does not yield meaningful limits -- regimes. In both cases, the favoured  muon $g-2$ region at the 2$\sigma$ level is shown in brown.

While here we have just focused on one scenario, we stress that it clearly illustrates the impact of our proposed search on the search for BSM physics, thus providing strong support to incorporate our proposal into the existing Exotic Higgs decays experimental programme.

\section{Conclusions}
\label{sec:conclusions}

Exotic decays of the 125 GeV Higgs boson provide a unique window into physics beyond the Standard Model, particularly for scenarios involving light new particles. In this work, we have introduced a novel search strategy for the exotic decay $h \to Z a$ in the $\ell\ell\tau\tau$ final state, where $a$ is a light pseudoscalar particle. Our analysis, performed at $\sqrt{s} = 14$ TeV and targeting the 300 fb$^{-1}$ and 3 ab$^{-1}$ datasets, probes 
pseudoscalar masses in the range $5 \leq m_a \leq 33$ GeV -- thus focusing on the kinematic region $m_h > m_Z + m_a$ where the exotic Higgs decay is two-body
-- .

Through a detailed validation of existing ATLAS and CMS searches for related Higgs decay channels, we have designed a robust selection strategy that optimizes sensitivity to $h \to Z a \to \ell\ell\tau\tau$. By leveraging both leptonic and hadronic tau decays, we have demonstrated that this $a\to \tau\tau$ channel provides a complementary and potentially more sensitive probe of light pseudoscalars compared to existing searches in $a \to \mu\mu$ and $a \to \gamma\gamma$ final states. We have established model-independent 95\% C.L. sensitivity projections 
on the branching ratio ${\rm BR}(h \to Z a) \times {\rm BR}(a \to \tau\tau)$ and compared them to the (expected) reach of the experimental searches currently performed and their corresponding future sensitivity projections. It is worth 
stressing that for a \textit{Standard-Model-like} Axion-Like Particle (ALP), our proposed search extends the expected limit on ${\rm BR}(h \to Z a)$ by almost two orders of magnitude w.r.t. existing search strategies. 

Additionally, we have explored the implications of this search in the context of ALP models motivated by the observed experimental anomaly in the value of the anomalous magnetic moment of the muon, $(g-2)_\mu$. We find that our proposed search could provide valuable constraints on the parameter space of a light ALP $a$ targeting $(g-2)_\mu$, in complementarity with existing $h\to Z a$, $a \to \gamma\gamma$ experimental searches.

Finally, given the projected increase in Higgs statistics at the HL-LHC, further refinements in tau-reconstruction techniques and machine learning-based event selection could significantly enhance the sensitivity of our proposed search. We encourage LHC experimental collaborations to incorporate this channel into future analyses, as it presents a promising avenue to probe new physics scenarios with light pseudoscalars, which can naturally have dominant decays into $\tau$ pairs.

\section*{Acknowledgements}

J.M.N. and J.Z. want to thank Nishita Desai for discussions that motivated us to think more seriously about this idea and which took place several years ago. We also thank Cecile Caillol for helpful comments on our analysis. J.M.N. and R.M.S.S.\ were supported by the Ram\'on y Cajal Fellowship contract RYC-2017-22986, and acknowledge partial financial support by the Spanish Research Agency (Agencia Estatal de Investigaci\'on) through the grants CNS2023-144536, PID2021-124704NB-I00, and IFT Centro de Excelencia Severo Ochoa No CEX2020-001007-S, funded by  
MCIN/AEI/10.13039/501100011033. J.M.N. also acknowledges partial financial support by the European Union's Horizon 2020 research and innovation programme under the Marie Sk\l odowska-Curie grant agreements No 860881-HIDDeN and 101086085-ASYMMETRY. J.Z is supported by the {\it Generalitat Valenciana} (Spain) through the {\it plan GenT} program (CIDEGENT/2019/068), by the Spanish Government (Agencia Estatal de
Investigaci\'on), ERDF funds from European Commission (MCIN/AEI/10.13039/501100011033, Grant No. PID2020-114473GB-I00 and No. PID2023-146220NB-I00), and by the Spanish Research Agency (Agencia Estatal
de Investigaci\'on, MCIU/AEI) through the grant IFIC Centro de Excelencia Severo Ochoa No. CEX2023-001292-S. M.C. acknowledges support by the Spanish Research Agency (Agencia Estatal de
Investigaci\'on) through the grants PID2023-147706NB-I00 and CNS2023-144781, funded by MCIU/AEI/10.13039/501100011033 and by European Union NextGenerationEU/PRTR.  
C.R.~is supported by 
Fundação Coordenação de Aperfeiçoamento de Pessoal de Nível Superior (CAPES) projects no.\ 88887.645500/2021-00 and 88887.935477/2024-00. 
Feynman diagrams were drawn using {\sc TikZ-Feynman}~\cite{Ellis:2016jkw}. The authors acknowledge support from the COMETA COST Action CA22130.

\bibliographystyle{JHEP}
\bibliography{HAZ_lltautau}

\end{document}